\RequirePackage[T1]{fontenc}\RequirePackage{fix-cm}\documentclass[letterpaper,journal]{IEEEtran}

\usepackage{amsmath,amsfonts,amssymb}
\usepackage{algorithmic}
\usepackage{algorithm}
\usepackage{array}
\usepackage{textcomp}
\usepackage{stfloats}
\usepackage{url}
\usepackage{verbatim}
\usepackage{graphicx}
\usepackage{placeins}
\usepackage{physics}
\usepackage{color}
\usepackage{bm}
\usepackage{subfigure}
\usepackage{booktabs}
\usepackage[dvipsnames]{xcolor}
\usepackage{cite}
\usepackage{balance}
\usepackage{setspace}
\usepackage{amsthm}
\usepackage{hyperref}

\AtBeginDocument{\raggedbottom}
\newcounter{ch23savedequation}
\newcounter{ch23savedalgorithm}
\newcounter{ch23equation}
\newcounter{ch23algorithm}

\allowdisplaybreaks[4]

\newtheoremstyle{compactproposition}
    {3pt}{3pt}{\itshape}{}{\bfseries}{.}{5pt}{}
\theoremstyle{compactproposition}
\newtheorem{thm}{Proposition}

\newtheorem{remark}{Remark}

\theoremstyle{definition}

\newtheorem*{pf}{Proof}

\newcommand{\algref}[1]{\textbf{Algorithm~\ref{#1}}}

\newcommand{\blu}[1]{\textcolor{black}{#1}}
\newcommand{\red}[1]{\textcolor{black}{#1}}
\newcommand{\gre}[1]{\textcolor{black}{#1}}
\definecolor{NarrativeOchre}{RGB}{0,0,0}
\definecolor{ComplexityTeal}{RGB}{0,0,0}
\definecolor{DeepPurple}{RGB}{160, 32, 240}

\floatname{algorithm}{Algorithm}

\renewcommand{\algorithmicdo}{\textbf{do}}

\renewcommand{\algorithmicthen}{\textbf{then}}

\DeclareMathAlphabet{\mathsfit}{\encodingdefault}{\sfdefault}{m}{sl}
\SetMathAlphabet{\mathsfit}{bold}{\encodingdefault}{\sfdefault}{bx}{n}
\newcommand{\tens}[1]{\bm{\mathsfit{#1}}}
\def\tA{{\tens{A}}}
\def\tB{{\tens{B}}}
\def\tC{{\tens{C}}}

\def\tF{{\tens{F}}}

\def\tK{{\tens{K}}}
\def\tL{{\tens{L}}}
\def\tM{{\tens{M}}}

\def\tQ{{\tens{Q}}}
\def\tR{{\tens{R}}}

\def\tU{{\tens{U}}}
\def\tV{{\tens{V}}}

\def\tX{{\tens{X}}}
\def\tY{{\tens{Y}}}
\def\tZ{{\tens{Z}}}

\definecolor{ReplacementGreen}{RGB}{0,0,0}

\makeatletter
\let\greenreview@float@makebox\float@makebox
\renewcommand{\float@makebox}[1]{%
  \color@vbox\normalcolor\greenreview@float@makebox{#1}\color@endbox%
}
\makeatother

\begin{document}

\bstctlcite{IEEEexample:BSTcontrol}

\title{Movable Antenna Aided Robust Wireless Transmission: From Model-Based Optimization to Scalable Reinforcement Learning}
    
\author{Jinshuo Zhang, \textit{Graduate Student Member, IEEE}, Yafei Wang, \textit{Graduate Student Member, IEEE},\\ Hongwei Hou, \textit{Graduate Student Member, IEEE}, Wenjin Wang, \textit{Member, IEEE}, Shi Jin,  \textit{Fellow}, \textit{IEEE},\\ Symeon Chatzinotas, \textit{Fellow}, \textit{IEEE}, Bj{\"o}rn Ottersten, \textit{Fellow}, \textit{IEEE}

\thanks{Jinshuo Zhang, Yafei Wang, Hongwei Hou and Wenjin Wang are with the National Mobile Communications Research Laboratory, Southeast University, Nanjing 210096, and also with Purple Mountain Laboratories, Nanjing 211100 (e-mail: zhangjs@seu.edu.cn; wangyf@seu.edu.cn; hongweihou@seu.edu.cn; wangwj@seu.edu.cn).}
\thanks{Shi Jin is with the National Mobile Communications Research Laboratory, Southeast University, Nanjing 210096 (jinshi@seu.edu.cn).}
\thanks{Symeon Chatzinotas and Bj{\"o}rn Ottersten are with the Interdisciplinary Centre for Security, Reliability and Trust (SnT), University of Luxembourg (E-mails: \{symeon.chatzinotas, bjorn.ottersten\}@uni.lu).}
}

    \markboth{}{Zhang \MakeLowercase{\textit{et al.}}: Movable Antenna-Aided Robust Wireless Transmission}

	\maketitle

        \begin{abstract}
        This paper investigates robust joint \textcolor{black}{antenna positioning} and precoding for multiuser movable antenna (MA) systems under imperfect channel state information (CSI). \blu{First, we identify the mismatch between the spatial statistics of densely sampled MA channels and the discrete Fourier transform (DFT) basis used for conventional robust precoding. To address this, we develop a novel discrete prolate spheroidal sequence (DPSS)-based \textit{a posteriori} channel model. By exploiting the spatial covariance structure of MA channels, this model is better suited to modeling MA channel uncertainty.} 
        \blu{Building on this model, we develop an efficient iterative algorithm for robust joint \textcolor{black}{antenna positioning} and precoding to maximize the expected sum rate. By transforming the coupled optimization of \textcolor{black}{antenna positioning} and precoding at each iteration into direct comparisons of closed-form  scores, we design a low-complexity position replacement scheme that effectively reuses matrix computations across candidate sets.} 
        \blu{To further improve efficiency and scalability, we exploit the inherent tensor equivariance (TE) of the problem to develop a reinforcement learning framework.} To explicitly account for channel uncertainty, we develop novel pairwise user-interaction and CSI-conditioned equivariant attention mechanisms to encode channel statistics and adaptively fuse them with the available CSI. We further introduce distance-biased spatial attention to incorporate the geometry of candidate positions into attention weighting, thereby enhancing spatial feature learning. \textcolor{black}{\textcolor{black}{Simulation results show that the proposed methods improve sum-rate performance while reducing online runtime under imperfect CSI. These methods also achieve near-optimal performance under perfect CSI, closely matching the exhaustive-search method.}} Moreover, the learning framework generalizes across different numbers of users and MAs without retraining.
	\end{abstract}


	\IEEEpeerreviewmaketitle

    \section{Introduction}
    
\IEEEPARstart{M}{ultiple-input} multiple-output (MIMO) technology is a key component of modern wireless communication systems, enabling substantial spectral-efficiency gains through spatial multiplexing and precoding~\cite{marzetta2010noncooperative,bjornson2014optimal}. \blu{Conventional MIMO systems typically employ fixed-position antenna (FPA) arrays, with precoding designed for channels associated with a fixed spatial configuration.} \blu{Recently, movable antennas (MAs) have emerged to break this constraint by enabling antenna moving within a prescribed region. By exploiting the additional spatial degrees of freedom, MAs can proactively reshape wireless channels to suppress inter-user interference and enhance spatial multiplexing, thereby achieving substantial spectral-efficiency gains~\cite{zhu2024modeling,zhu2024opportunities,wan2026toward}.}

\blu{To fully exploit the potential of MAs in multiuser transmission, existing studies have explored joint antenna positioning and precoding designs, which can be broadly categorized into continuous and discrete positioning paradigms. For continuous positioning, \cite{feng2024weightedsumrate} combined weighted minimum mean square error (WMMSE) beamforming with majorization--minimization position updates for weighted sum-rate maximization. Joint antenna positioning and regularized zero-forcing (RZF) precoding were addressed through regularized sparse optimization in~\cite{yang2024flexible}. While continuous positioning offers substantial performance gains, the finite positioning resolution of practical MA actuators has motivated research on discrete positioning, where solving the resulting combinatorial problems optimally may incur exorbitant computational complexity. To address this challenge, \cite{wu2023discrete} used generalized Benders decomposition to minimize transmit power under user signal-to-interference-plus-noise ratio (SINR) constraints. Beyond power minimization, \cite{yang2025flexiblewmmse} developed a flexible WMMSE framework for weighted sum-rate maximization through regularized sparse optimization over predefined position dictionaries.}

\blu{While the above methods achieve substantial performance gains, they typically assume perfect channel state information (CSI). In practice, however, channel uncertainty is unavoidable and can decrease the performance gains achieved through both antenna positioning and precoding~\cite{6608213,9416909}. This has motivated research on robust joint positioning and precoding designs for MA systems.} The two-timescale design in \cite{zheng2025twotimescale} optimizes antenna positions using statistical CSI and adapts precoding to instantaneous CSI. For single-user systems, \cite{ma2025robust} combines maximum-ratio transmission with graph-based antenna positioning under norm-bounded and Gaussian CSI error models. For multiuser systems, \cite{wu2025globallyoptimal} jointly optimizes discrete antenna positions and beamforming under norm-bounded errors in the path-response coefficients to minimize total \textcolor{black}{base station (BS)} power while satisfying worst-case user SINR constraints. \blu{In addition, \textit{a posteriori} channel models have been adopted in conventional massive MIMO systems to capture channel variations and spatial correlation, enabling robust precoding to adapt to changing channel conditions and improve spectral efficiency~\cite{8694866,shi2021dlrobust}, offering a promising basis for channel uncertainty modeling in MA systems.}

\blu{To improve transmission performance while reducing online computational overhead, various deep learning methods have been investigated in MA systems.} For two-timescale design, deep unfolding accelerates joint optimization of antenna positions based on long-term channel statistics and beamforming based on short-term CSI~\cite{liang2026twotimescaledeep}. Under imperfect CSI, \cite{weng2024learningma} models channel estimation errors as correlated complex Gaussian perturbations and uses reinforcement learning (RL) to jointly optimize transmit beamforming and antenna movement at the transmitter and receivers. Moreover, permutation equivariance/invariance (PE/PI) has been exploited in a \textcolor{black}{graph neural network (GNN)}-based framework for joint \textcolor{black}{antenna positioning} and precoding in MA systems under perfect CSI~\cite{wang2026learningjoint}. More generally, the tensor equivariance (TE) framework extends these properties to multidimensional and high-order forms for conventional UPA systems, enabling low-complexity precoding and generalization across different system sizes without retraining~\cite{11049893,zhang2025unlocking,zhang2026accelerate}, highlighting its potential for robust joint \textcolor{black}{antenna positioning} and precoding in MA systems.

\blu{Although existing works have incorporated estimation and feedback errors into MA transmission design~\cite{wu2025globallyoptimal,weng2024learningma}, accurately modeling channel uncertainty for densely sampled MAs remains a major challenge. Conventional DFT-based diagonal covariance representations impose a periodic structure that may not match the spatial correlation of densely sampled MA channels, resulting in a structural covariance mismatch. Consequently, developing low-complexity robust joint \textcolor{black}{antenna positioning} and precoding methods tailored to these spatial characteristics remains an urgent problem to be addressed.} Moreover, although equivariance has been exploited in GNN-based MA designs under perfect CSI~\cite{wang2026learningjoint}, the TE properties of robust joint \textcolor{black}{antenna positioning} and precoding and their potential for generalization across various numbers of users and MAs remain insufficiently explored. These considerations motivate the following question: \textit{\blu{How can we achieve efficient and scalable robust transmission design for multiuser MA systems under imperfect CSI?}} To address this question, we then develop a robust joint \textcolor{black}{antenna positioning} and precoding framework combining statistical channel modeling, and a scalable reinforcement learning method. The main contributions are as follows.
\begin{itemize}
    \item We first investigate channel uncertainty modeling across densely spaced MA candidate positions and identify a structural mismatch of the discrete Fourier transform (DFT)-based covariance representation commonly used in conventional robust precoding. To better capture the spatial characteristics, we employ a discrete prolate spheroidal sequence (DPSS) basis to establish an \textit{a posteriori} channel model over all candidate positions. The DPSS basis is constructed from a finite-aperture reference covariance and combined with user-specific long-term statistics to characterize the covariance of channel uncertainty. Building on this channel model, we establish the robust multiuser MA system model and formulate the joint \textcolor{black}{antenna positioning} and precoding problem to maximize the expected sum rate under imperfect CSI.
    \item For the formulated problem, we propose the \textcolor{black}{robust position-adaptive WMMSE (RPA-WMMSE)} algorithm based on scoring of candidate position combinations. \textcolor{black}{We derive a closed-form score that incorporates spatially correlated posterior channel uncertainty and accounts for precoding adaptation, enabling direct comparison among feasible candidate position combinations.} We further exploit the common structure of combinations that differ in a single position to decompose their scores into a shared term and candidate-dependent increments. This decomposition enables efficient evaluation of feasible replacements by reusing the matrix factorization associated with the retained positions.
    \item To further improve online computational efficiency, we exploit the TE structure of the joint design problem to develop \textcolor{black}{PSPCNet, a model-driven reinforcement learning framework for robust joint \textcolor{black}{antenna positioning} and precoding}. To explicitly account for channel uncertainty, we introduce pairwise user-interaction encoding that captures both multiuser interference and features in channel statistics. We further design CSI-conditioned equivariant attention to adaptively fuse the covariance features of selected positions with the available CSI for robust precoding. We also introduce distance-biased spatial attention to explicitly incorporate the physical geometry of candidate positions into attention weighting for position selection. Equivariant feature processing and a decoder with shared parameters across position-selection steps enable the same trained model to support different numbers of users and MAs without retraining.
\end{itemize}

\textcolor{black}{This paper is organized as follows: The system model and problem formulation are presented in Section~\ref{Section 2}. The RPA-WMMSE algorithm is developed in Section~\ref{sec:ao-robust}, followed by the model-driven reinforcement learning framework in Section~\ref{sec:rl-framework}. Section~\ref{Section 6} reports simulation results, and Section~\ref{Section 7} concludes the paper.}

\textit{Notation:} Scalars, column vectors, matrices, and tensors are denoted by $x$, $\mathbf x$, $\mathbf X$, and $\tX$, respectively.
$\mathbb R^{M\times N}$ and $\mathbb C^{M\times N}$ denote the spaces of real and complex $M\times N$ matrices, respectively.
The operations of transpose, complex conjugation, conjugate transpose, and matrix inversion are denoted by $(\cdot)^T$, $(\cdot)^*$, $(\cdot)^H$, and $(\cdot)^{-1}$, respectively.
$\|\cdot\|_2$ and $\|\cdot\|_{\rm F}$ denote the Euclidean and Frobenius norms, respectively.
$\operatorname{Tr}(\cdot)$ denotes the matrix trace, and $\operatorname{diag}(\mathbf x)$ forms a diagonal matrix with $\mathbf x$ on its diagonal.
$\otimes$ and $\odot$ denote the Kronecker and Hadamard products, respectively.
$\mathfrak R\{\cdot\}$ and $\mathfrak I\{\cdot\}$ extract the real and imaginary parts, respectively.
$|\cdot|$ denotes magnitude and acts elementwise on arrays.
$[\mathbf X]_{i,j}$ and $[\mathbf x]_i$ denote the corresponding matrix and vector entries.
$[\cdot]_D$ denotes concatenation along the existing $D$-th dimension, whereas $[\cdot]_{(D)}$ denotes stacking along a newly inserted $D$-th dimension.
$\mathbb E\{\cdot\}$ denotes expectation, and $\mathcal{CN}(0,\sigma^2)$ denotes a circularly symmetric complex Gaussian distribution with zero mean and variance $\sigma^2$.

\section[MA Channel Uncertainty Modeling and Problem Formulation]{\textcolor{black}{MA Channel Uncertainty Modeling and Problem Formulation}}\label{Section 2}
\subsection{Signal Model}\label{system model}

\begin{figure}[!t]
    \centering
    \includegraphics[width=0.9\columnwidth]{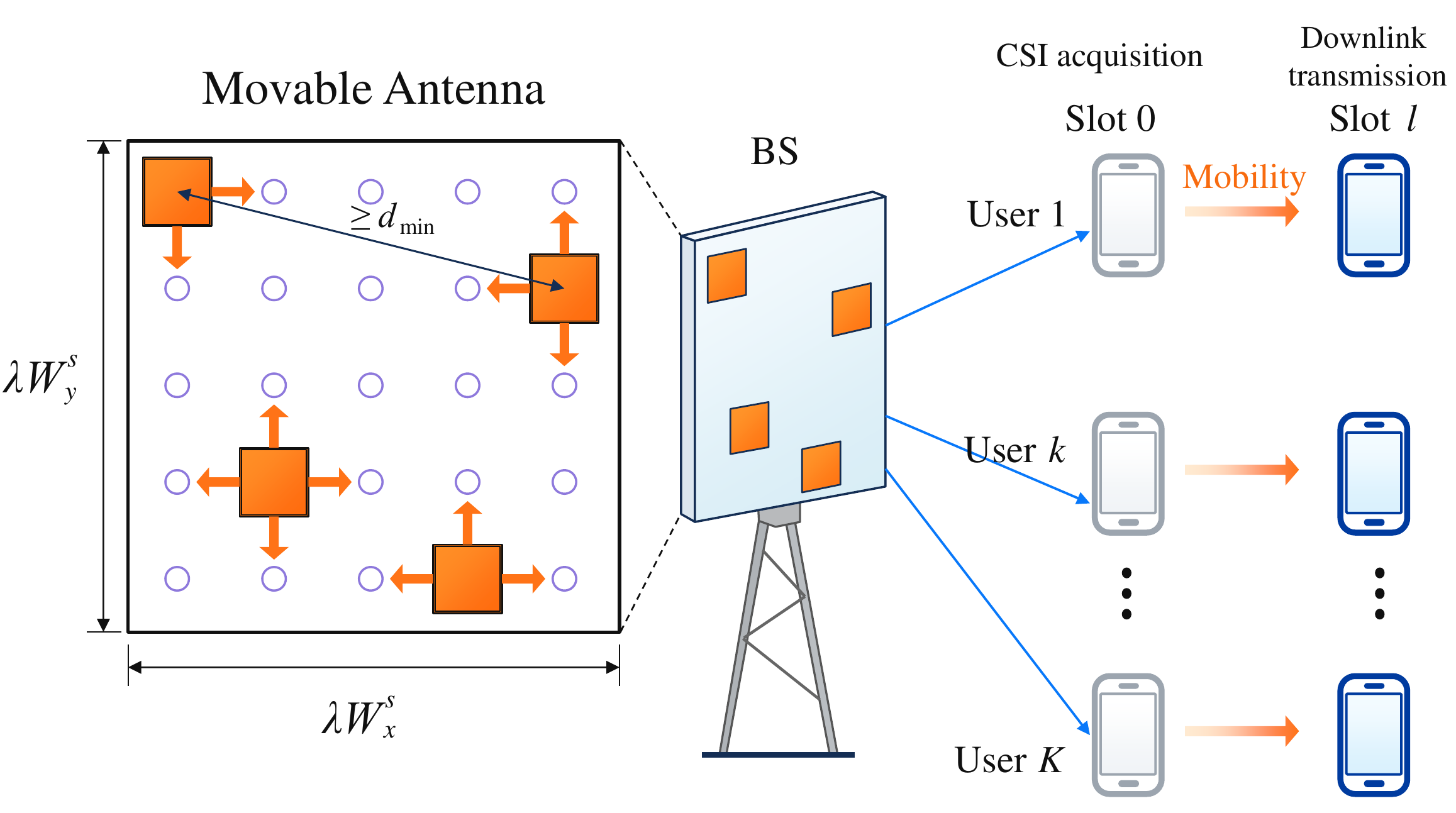}
    \caption{MA downlink with \textcolor{black}{antenna positioning at the BS} for multiple mobile users.}
    \label{fig:fas-system-model}
    \vspace{-16pt}
\end{figure}

\textcolor{black}{As shown in Fig.~\ref{fig:fas-system-model}, we consider a time-division duplex (TDD) MIMO downlink system in which a \textcolor{black}{BS} equipped with $n_{\rm s}$ MAs serves $K$ single-antenna users.} \textcolor{black}{The positions of the MAs can be individually adjusted within a rectangular moving region of size $\lambda W_x^{\rm s}\times\lambda W_y^{\rm s}$, where $\lambda$ is the wavelength and $W_x^{\rm s}$ and $W_y^{\rm s}$ are the normalized side lengths along the $x$ and $y$ axes, respectively.}
 \textcolor{black}{\textcolor{black}{We consider discrete antenna positioning, where the} $n_{\rm s}$ MAs occupy distinct positions on a uniform grid of $N_{\rm s}=N_x^{\rm s}\times N_y^{\rm s}$ candidate positions within this region, with $K\le n_{\rm s}\le N_{\rm s}$ and pairwise distances of at least $d_{\min}$~\cite{wu2025globallyoptimal}.} \textcolor{black}{Let $\mathbf r_n=[x_n,y_n]^T\in\mathbb R^{2\times1}$ denote the coordinates of candidate position $n$, for $n=1,\ldots,N_{\rm s}$.}
\textcolor{black}{Channel estimates obtained from uplink training are used for downlink transmission over the subsequent $L$ slots.}

Let $x_k[l]$ denote the symbol transmitted in slot $l$ to user $k$, where $l\in\mathcal L=\{1,2,\ldots,L\}$ and $k\in\mathcal K=\{1,2,\ldots,K\}$. \textcolor{black}{The position of the $i$-th MA in slot $l$ is $\mathbf r_{m_i[l]}$, where $m_i[l]\in\{1,\ldots,N_{\rm s}\}$ denotes its selected candidate-position index.} The received signal at user $k$ is
\begin{equation}
    y_k[l]=\mathbf h_k^H[l]\mathbf E[l]\mathbf P[l]\mathbf x[l]+n_k[l],
    \label{rev23:eq:received-signal}
\end{equation}
\textcolor{black}{where $\mathbf x[l]=[x_1[l],\ldots,x_K[l]]^T\in\mathbb C^{K\times1}$ is the zero-mean transmitted symbol vector satisfying $\mathbb E\{\mathbf x[l]\mathbf x^H[l]\}=\mathbf I_K$, $\mathbf h_k[l]\in\mathbb C^{N_{\rm s}\times1}$ is the channel vector from all $N_{\rm s}$ candidate positions to user $k$, $\mathbf E[l]\triangleq[\mathbf e_{m_1[l]},\ldots,\mathbf e_{m_{n_{\rm s}}[l]}]\in\mathbb R^{N_{\rm s}\times n_{\rm s}}$ is the position selection matrix with $\mathbf e_n$ denoting the $n$-th column of $\mathbf I_{N_{\rm s}}$, $\mathbf P[l]\in\mathbb C^{n_{\rm s}\times K}$ is the precoding matrix, and $n_k[l]\sim\mathcal{CN}(0,\sigma^2)$ is the additive Gaussian noise with power $\sigma^2$. The vector $\mathbf E^T[l]\mathbf h_k[l]$ contains the entries of $\mathbf h_k[l]$ corresponding to the $n_{\rm s}$ selected MA positions.}

\subsection[The A Posteriori Channel Model for MA Systems]{\textcolor{black}{The A Posteriori Channel Model for MA Systems}} \label{sec:CIR}

\textcolor{black}{To characterize the CSI uncertainty arising from channel variations between uplink estimation and downlink transmission, we adopt an \textit{a posteriori} channel model~\cite{8694866,wang2024robust}.} Let $\mathbf h_k[0]$ denote the channels from all $N_{\rm s}$ positions to user $k$ estimated by the BS during the uplink phase. The channel over all candidate positions in slot $l$ is then expressed as
\begin{equation}
    \mathbf h_k[l]=\alpha_k[l]\mathbf h_k[0]
    +\beta_k[l]\mathbf V\big(\mathbf m_k\odot\mathbf g_k[l]\big),
    \label{rev23:eq:posterior-channel}
\end{equation}
where $\alpha_k[l]$ is the temporal correlation coefficient of the channel to user $k$ in slot $l$, which can be characterized by the Jakes autocorrelation model~\cite{jakes1974microwave,6608213}, and $\beta_k[l]=\sqrt{1-\alpha_k^2[l]}$. The vector $\mathbf m_k\in\mathbb R^{N_{\rm s}\times1}$ is a nonnegative statistical amplitude vector in the transform domain that remains unchanged over the long term, and the elements of $\mathbf g_k[l]\in\mathbb C^{N_{\rm s}\times1}$ are independently and identically distributed as $\mathcal{CN}(0,1)$. \blu{$\mathbf V\in\mathbb C^{N_{\rm s}\times N_{\rm s}}$ is the spatial sampling matrix used to characterize the spatial correlation of the channel, commonly represented by a DFT matrix for uniform linear arrays with a large number of antennas. However, this DFT-based representation may be inaccurate for the densely spaced MA candidate positions considered in this work, and we therefore construct $\mathbf V$ using DPSS in the following subsection.} Given the available CSI, this model yields the channel mean over all candidate positions $\overline{\mathbf h}_k[l]=\alpha_k[l]\mathbf h_k[0]$ and covariance $\mathbf R_k[l]=\beta_k^2[l]\mathbf V\operatorname{diag}(|\mathbf m_k|^2)\mathbf V^H$.

\textcolor{black}{We define the MA position configuration in slot $l$ as the ordered tuple $\mathcal S[l]=(m_1[l],\ldots,m_{n_{\rm s}}[l])$.} Let $\mathbf h_{k,\mathcal S}[l]\in\mathbb C^{n_{\rm s}\times1}$ denote the channel in slot $l$ from the selected positions to user $k$. We then have
\begin{equation}
\begin{aligned}
    \mathbf h_{k,\mathcal S}[l]
    =\alpha_k[l]\mathbf E^T[l]\mathbf h_k[0]
    +\beta_k[l]\mathbf E^T[l]\mathbf V\big(\mathbf m_k\odot\mathbf g_k[l]\big).
\end{aligned}
    \label{rev23:eq:selected-port-channel}
\end{equation}
The corresponding selected-position channel has mean $\overline{\mathbf h}_{k,\mathcal S}[l]=\mathbf E^T[l]\overline{\mathbf h}_k[l]$ and covariance $\mathbf R_{k,\mathcal S}[l]=\mathbf E^T[l]\mathbf R_k[l]\mathbf E[l]$.

\subsection[Spatial Covariance Modeling for MA Systems]{\textcolor{black}{Spatial Covariance Modeling for MA Systems}}\label{subsec:ma-spatial-covariance}

\begin{figure}[t]
    \centering
    \includegraphics[width=0.9\columnwidth]{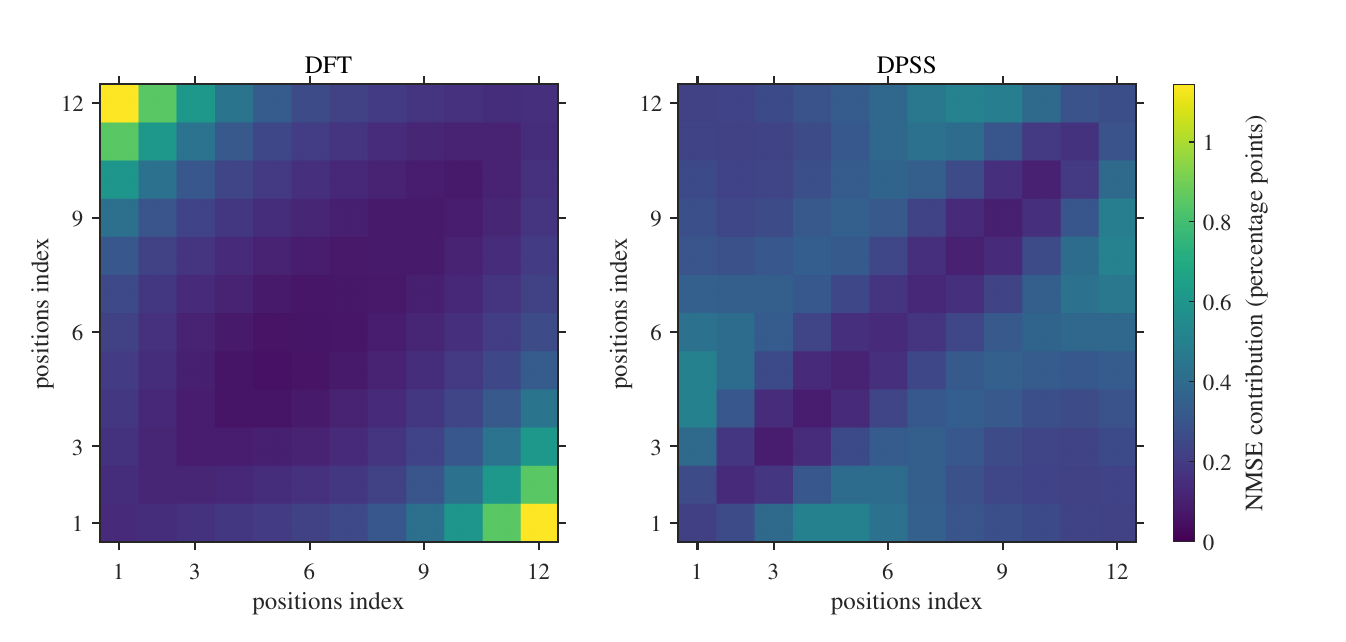}
    \caption{\textcolor{black}{Normalized MSE distribution of covariance approximations using DFT and DPSS bases ($N=12$, $\delta=1/8$).}}
    \label{fig:dft-dpss-covariance-error}
    \vspace{-16pt}
\end{figure}

\red{%
We next analyze the spatial statistics of MA channels to construct a suitable transform basis $\mathbf V$. The channel modeled by~\eqref{rev23:eq:posterior-channel} has covariance
\begin{equation}
    \mathbf R_k[l]
    =\beta_k^2[l]\mathbf V
    \operatorname{diag}\!\left(|\mathbf m_k|^2\right)\mathbf V^H,
    \label{dpss-review:eq:posterior-covariance}
\end{equation}
where $\mathbf V$ determines the covariance structures that this model can represent. Therefore, selecting a transform basis that matches the spatial statistics of MA channels is essential.
}

\begingroup\color{black}
\begin{remark}\label{dpss-review:remark:sampling}
When $\mathbf V$ is chosen as the DFT basis commonly used for UPAs, the model imposes a circulant structure on $\mathbf R_k[l]$, whereas spatially stationary channels sampled on a uniform grid generally have a Toeplitz covariance that does not satisfy this periodicity~\cite{gray2006toeplitz}. For UPAs, the DFT approximation becomes asymptotically accurate as the aperture grows at fixed antenna spacing~\cite{lu2022robust3d,you2015pilotreuse}. However, for MA candidates within a fixed aperture, this periodic constraint can introduce a structural covariance mismatch.
\end{remark}
\endgroup

To characterize the spatial statistics of MA channels, we consider $N$ uniformly spaced candidate positions over a one-dimensional moving region of length $\lambda W$, with normalized spacing $\delta=W/(N-1)\in(0,1/2)$. Under far-field propagation~\cite{zhu2024modeling}, the channel at position $n$ can be expressed as
\begin{equation}
    h_n=\sum_{p=1}^{N_{\mathrm p}}c_p e^{j2\pi(n-1)\delta\sin\theta_p},
    \qquad n=1,\ldots,N,
    \label{dpss-review:eq:far-field-channel}
\end{equation}
where $N_{\mathrm p}$ is the number of propagation paths, and $c_p$ and $\theta_p$ denote the complex gain and AoD of path $p$, respectively. Each path has normalized spatial frequency $f_p=\delta\sin\theta_p\in[-\delta,\delta]$.

To construct a common transform basis from this spatial-frequency support, we consider a reference model with uncorrelated scattering components and unit spatial power spectral density over $[-\delta,\delta]$. This reference captures the spatial-frequency support determined by the candidate spacing without specifying a user-dependent power distribution. Its covariance is denoted by $\mathbf R_{\rm ref}$.

\begingroup\color{black}
\begin{thm}\label{dpss-review:prop:dpss}
Under the above reference model, the covariance across candidate positions is
\begin{equation}
\begin{aligned}
    [\mathbf R_{\rm ref}]_{m,n}\!
    =\!\int_{-\delta}^{\delta}e^{j2\pi f(m-n)}\,\mathrm df
    \!=\!2\delta\,\operatorname{sinc}\!\left(2\delta(m-n)\right),
\end{aligned}
\label{dpss-review:eq:reference-covariance}
\end{equation}
for $m,n=1,\ldots,N$, where $\operatorname{sinc}(x)=\sin(\pi x)/(\pi x)$, with $\operatorname{sinc}(0)=1$.
This covariance is a finite prolate matrix, whose orthonormal eigenvectors form the classical DPSS basis $\mathbf U_{\rm D}$ of length $N$ and normalized half-bandwidth $\delta$~\cite{slepian1978dpss,karnik2017fastslepian}. Thus,
\begin{equation}
    \mathbf R_{\rm ref}
    =\mathbf U_{\rm D}\boldsymbol\Lambda_{\rm D}\mathbf U_{\rm D}^H,
    \label{dpss-review:eq:dpss-decomposition}
\end{equation}
where $\boldsymbol\Lambda_{\rm D}$ is the diagonal matrix of the eigenvalues.
\end{thm}

\begin{pf}
\textit{See Appendix~\ref{app:proof-dpss-reference}.}
\end{pf}
\endgroup

We therefore adopt $\mathbf U_{\rm D}$ as the transform basis for the one-dimensional statistical MA channel representation.
Fig.~\ref{fig:dft-dpss-covariance-error} compares the entrywise errors of DFT- and DPSS-based diagonal approximations to sample spatial correlation matrices. As shown, the DFT error is pronounced for position pairs at opposite grid endpoints due to the structural mismatch introduced by the circulant covariance constraint.

For the two-dimensional rectangular moving region considered in this work, we construct one-dimensional DPSS bases $U_{{\rm D},x}$ and $U_{{\rm D},y}$ along the $x$ and $y$ axes, and
their Kronecker product is adopted as the spatial statistical transform basis~\cite{salvorossi2008slepian2d}
\begin{equation}
    {\mathbf U}_{\rm 2D}
    =\mathbf U_{{\rm D},y}\otimes\mathbf U_{{\rm D},x}
    \in\mathbb R^{N_{\rm s}\times N_{\rm s}}.
    \label{dpss-review:eq:two-dimensional-basis}
\end{equation}
This construction provides a common statistical basis determined by the candidate-grid geometry for general MA channels. Substituting this DPSS basis for $\mathbf V$ in~\eqref{rev23:eq:posterior-channel} yields the \textit{a posteriori} channel model used to account for spatially correlated uncertainty in robust joint antenna positioning and precoding.


\subsection[Robust Joint Antenna Positioning and Precoding]{\textcolor{black}{Robust Joint \textcolor{black}{Antenna Positioning} and Precoding}}\label{sec:problem-formulation}

For MA downlink transmission, the \textcolor{black}{SINR} of user $k$ in slot $l$ is
\begin{equation}
    \gamma_k[l]=
    \frac{|\mathbf h_k^H[l]\mathbf E[l]\mathbf p_k[l]|^2}
    {\sum_{j\ne k}|\mathbf h_k^H[l]\mathbf E[l]\mathbf p_j[l]|^2+\sigma^2},
    \label{eq:instantaneous-sinr}
\end{equation}
where $\mathbf p_k[l]\in\mathbb C^{n_{\rm s}\times1}$ is the column of $\mathbf P[l]$ corresponding to user $k$. Let $\mathbf H[l]=[\mathbf h_1[l],\ldots,\mathbf h_K[l]]^H\in\mathbb C^{K\times N_{\rm s}}$ denote the channel matrix over all candidate positions. Under imperfect CSI, the expected rate of user $k$ is
\begin{equation}
    \mathcal R_k(\mathbf H[l],\mathbf E[l],\mathbf P[l],\sigma^2)
    =\mathbb E\!\left\{\log\big(1+\gamma_k[l]\big)\right\},
    \label{eq:objective function}
\end{equation}
where the expectation is taken over $\mathbf H[l]$. The robust joint \textcolor{black}{antenna positioning} and precoding problem is formulated as
\begingroup\color{black}
\begin{equation}
\begin{aligned}
    \max_{\mathbf E[l],\mathbf P[l]}\quad
    &\sum_{k=1}^K\mathcal R_k(\mathbf H[l],\mathbf E[l],\mathbf P[l],\sigma^2)\\
    \text{s.t.}\quad
    &\mathbf E[l]\in\{0,1\}^{N_{\rm s}\times n_{\rm s}},\\
    &\mathbf E^T[l]\mathbf 1_{N_{\rm s}}=\mathbf 1_{n_{\rm s}},\\
    &{\|\mathbf r_{m_i[l]}-\mathbf r_{m_j[l]}\|_2\ge d_{\min},
      \quad 1\le i<j\le n_{\rm s},}\\
    &\|\mathbf P[l]\|_{\rm F}^2\le P_{\rm T},
\end{aligned}
\label{eq:problem wmmse 1}
\end{equation}
\endgroup
where $P_{\rm T}$ is the maximum transmit power. The discrete position selection variables and the nonconcavity of the
expected sum rate with respect to the precoding matrix lead to a
nonconvex combinatorial problem, making a globally optimal solution
difficult to obtain.

\section[Robust Joint Antenna Positioning and Precoding via Closed-Form Scoring]{\textcolor{black}{Robust Joint \textcolor{black}{Antenna Positioning} and Precoding via Closed-Form Scoring}}\label{sec:ao-robust}

\begingroup\color{black}
In this section, we develop \textcolor{black}{the RPA-WMMSE} algorithm for robust joint \textcolor{black}{antenna positioning} and precoding. To avoid repeated precoding optimization during position selection, we derive closed-form scores that incorporate the corresponding robust precoding solution, enabling direct comparison of candidate position sets. To further reduce the computational complexity, we exploit the matrix blocks shared by single-position replacements to derive incremental scores that can be evaluated using a common matrix factorization.

\endgroup

    \subsection[Robust WMMSE Formulation for Joint Antenna Positioning and Precoding]{\textcolor{black}{Robust WMMSE Formulation for Joint \textcolor{black}{Antenna Positioning} and Precoding}}\label{sec:lower-bound}
    
    Since the objective function involves expectation, the joint optimization problem~\eqref{eq:problem wmmse 1} is difficult to solve directly. To this end, an expected sum-rate lower bound is constructed using the posterior channel model to simplify the subsequent optimization. In the following, the slot index $l$ is omitted, and we denote $\overline{\mathbf h}_k=\mathbb E\{\mathbf h_k\}$. For the selected channel $\mathbf E^T\mathbf h_k$, we derive the lower bound on the expected rate
of user $k$~\cite{shi2023robustwmmse}:
    \begin{equation}
    \begin{aligned}
        &\mathcal R_k^{\rm lb}(\mathbf H,\mathbf E,\mathbf P,\sigma^2)
        =\log\!\left(
        \sigma^2+\sum_{i=1}^{K}\mathbb E\!\left\{
        \left|\mathbf h_k^H\mathbf E\mathbf p_i\right|^2
        \right\}\right)\\
        &\quad-\log\!\left(
        \sigma^2+\sum_{i=1}^{K}\mathbb E\!\left\{
        \left|\mathbf h_k^H\mathbf E\mathbf p_i\right|^2
        \right\}
        -\left|\overline{\mathbf h}_k^H\mathbf E\mathbf p_k\right|^2
        \right).
    \end{aligned}
    \label{eq:lower bound}
    \end{equation}
    \begin{align}
        \mathbb E\!\left\{\!|\mathbf h_k^H\mathbf E\mathbf p_j|^2\!\right\}
        \!=\!\left|\overline{\mathbf h}_k^H\mathbf E\mathbf p_j\right|^2\!\!+\!\beta_k^2\mathbf p_j^H\mathbf E^T\mathbf V\boldsymbol{\Psi}_k\mathbf V^H\mathbf E\mathbf p_j,\label{eq:expectation term 2}
    \end{align}
    where $\boldsymbol{\Psi}_k=(\mathbf m_k\mathbf m_k^T)\odot\mathbf I_{N_{\rm s}}$. When CSI is perfectly known, i.e., $\mathbf h_k=\overline{\mathbf h}_k$, this lower bound coincides with the expected rate in~\eqref{eq:objective function}. We then adopt the sum of the user rate lower bounds as the objective
for joint \textcolor{black}{antenna positioning} and precoding, and reformulate its
maximization as an equivalent robust WMMSE problem.
Let $\boldsymbol{\mu}=[\mu_1,\ldots,\mu_K]^T
\in\mathbb C^{K\times1}$ denote the receiver coefficient vector,
with $\mu_k^*y_k$ representing the symbol estimate at user $k$.
The corresponding mean square error (MSE) is given by
\begin{equation}
\begin{aligned}
\mathcal E_k
&=\mathbb E\!\left\{\left|\mu_k^*y_k-x_k\right|^2\right\}\\
&=1-\mu_k^*\overline{\mathbf h}_k^H\mathbf E\mathbf p_k
-\mathbf p_k^H\mathbf E^T\overline{\mathbf h}_k\mu_k\\
&\quad+|\mu_k|^2\!\left(
\sum_{j=1}^{K}\mathbb E\!\left\{
\mathbf h_k^H\mathbf E\mathbf p_j\mathbf p_j^H
\mathbf E^T\mathbf h_k
\right\}+\sigma^2\right),
\end{aligned}
\label{eq:robust-mse}
\end{equation}
where $\mathbb E\!\left\{
\mathbf h_k^H\mathbf E\mathbf p_j\mathbf p_j^H
\mathbf E^T\mathbf h_k
\right\}$ can be evaluated using~\eqref{eq:expectation term 2}. With the weight coefficient vector
$\boldsymbol{\nu}=[\nu_1,\ldots,\nu_K]^T
\in\mathbb C^{K\times1}$,
the resulting robust WMMSE problem is expressed
as~\cite{shi2023robustwmmse,christensen2008wmmse,shi2011wmmse}:
\begingroup\color{black}
\interdisplaylinepenalty=10000
\begin{equation}
\begin{aligned}
\min_{\boldsymbol{\mu},\boldsymbol{\nu},\mathbf E,\mathbf P}\quad
&\sum_{k=1}^{K}\left(\nu_k\mathcal E_k-\log\nu_k\right)\\
\text{s.t.}\quad
&\mathbf E\in\{0,1\}^{N_{\rm s}\times n_{\rm s}},\\
&\mathbf E^T\mathbf 1_{N_{\rm s}}=\mathbf 1_{n_{\rm s}},\\
&{\|\mathbf r_{m_i}-\mathbf r_{m_j}\|_2\ge d_{\min},
\quad 1\le i<j\le n_{\rm s},}\\
&\operatorname{Tr}(\mathbf P\mathbf P^H)\le P_{\rm T}.
\end{aligned}
\label{eq:problem wmmse 2}
\end{equation}
\endgroup
This reformulation replaces the logarithmic rate expressions
with a more tractable weighted MSE objective. However, the joint
optimization of discrete \textcolor{black}{antenna positioning} and continuous precoding
remains a nonconvex combinatorial problem. \textcolor{black}{To solve this problem efficiently, the following subsections develop a position replacement algorithm based on closed-form position-set scores.}

\begingroup\color{black}
\begingroup

\red{\subsection{Joint \textcolor{black}{Antenna Positioning} and Precoding via Closed-Form Incremental Scoring}\label{subsec:position-score}\label{subsec:position-replacement}}

\blu{We then develop an efficient robust position-adaptive WMMSE (RPA-WMMSE) algorithm for robust joint \textcolor{black}{antenna positioning} and precoding. Specifically,  the proposed algorithm alternately updates the variable pairs $(\boldsymbol\mu,\boldsymbol\nu)$ and $(\mathbf E,\mathbf P)$. To maintain low computational complexity during this alternating procedure, we derive a closed-form position-set score that enables the direct comparison of feasible candidate sets.}

\red{\blu{For fixed $\mathbf E$ and $\mathbf P$, the receiver coefficients $\boldsymbol\mu$ and weights $\boldsymbol\nu$ are updated as~\cite{shi2011wmmse,christensen2008wmmse,shi2023robustwmmse}}
\begin{align}
    \mu_k
    &=\left(\sigma^2+\sum_{j=1}^{K}
    \mathbb E\!\left\{|\mathbf h_k^H\blu{\mathbf E}\blu{\mathbf p_j}|^2\right\}\right)^{-1}
    \overline{\mathbf h}_k^H\blu{\mathbf E}\blu{\mathbf p_k},
    \label{Receiver}\\
    \nu_k
    &=\left(1-\mu_k^*
    \overline{\mathbf h}_k^H\blu{\mathbf E}\blu{\mathbf p_k}\right)^{-1},
    \label{Weight}
\end{align}
\blu{where $\mathbf p_k$ is the $k$-th column of $\mathbf P$, and the expectation $\mathbb E\!\left\{|\mathbf h_k^H\mathbf E\mathbf p_j|^2\right\}$ is evaluated using~\eqref{eq:expectation term 2}.}}

\begingroup
\blu{With $\boldsymbol\mu$ and $\boldsymbol\nu$ fixed, we jointly optimize $\mathbf E$ and $\mathbf P$ in~\eqref{eq:problem wmmse 2}.
This problem remains combinatorial due to the discrete position selection, and solving it directly incurs a high computational cost. To address this challenge, we next derive a closed-form position-set score to directly compare feasible position sets. To further reduce computational complexity, we develop a sequential position-replacement method based on a computationally efficient incremental form of this score.}

\endgroup

\subsubsection{Closed-Form Position-Set Scoring}
\begingroup\leavevmode\color{black}
\blu{By incorporating the sum-power constraint into the noise term of the robust MSE in~\eqref{eq:robust-mse}~\cite{hu2021deepunfolding,yang2025flexiblewmmse},} the objective in~\blu{\eqref{eq:problem wmmse 2}} can be equivalently written as
\begin{equation}
    \sum_{k=1}^{K}\left(\nu_k\widetilde{\mathcal E}_k-\log\nu_k\right)=\sum_{k=1}^{K}(\nu_k-\log\nu_k)+f(\mathbf E,\mathbf P),
    \label{eq:robust-weighted-mse-objective}
\end{equation}
where $\widetilde{\mathcal E}_k$ is obtained from the MSE in~\eqref{eq:robust-mse} by replacing $\sigma^2$ with $\sigma^2\|\mathbf P\|_{\mathrm F}^2/P_{\mathrm T}$. Since $\sum_{k=1}^{K}(\nu_k-\log\nu_k)$ is independent of $\mathbf E$ and $\mathbf P$, minimizing the reformulated objective reduces to minimizing $f(\mathbf E,\mathbf P)$, given by
\begin{align}
    \mbox{\small$\displaystyle f(\mathbf E,\mathbf P)=$}
    &\mbox{\small$\displaystyle\sum_{k=1}^{K}\nu_k\Bigg[
    |\mu_k|^2\Bigg(
    \sum_{j=1}^{K}\mathbb E\!\left\{
    \left|\mathbf h_k^H\mathbf E\mathbf p_j\right|^2
    \right\}+\frac{\sigma^2}{P_{\rm T}}
    \|\mathbf P\|_{\rm F}^2\Bigg)$} \notag\\
    &\mbox{\small$\displaystyle{}-2\operatorname{Re}\!\left\{
    \mu_k^*\overline{\mathbf h}_k^H\mathbf E\mathbf p_k
    \right\}\Bigg].$} \label{eq:profiled-quadratic}
\end{align}
To express this robust objective in matrix form, we first define the channel correlation matrices $\mathbf Q_k\in\mathbb C^{N_{\rm s}\times N_{\rm s}}$ over all candidate positions as
\begin{equation}
    \mathbf Q_k\triangleq\mathbb E\{\mathbf h_k\mathbf h_k^H\}
    =\overline{\mathbf h}_k\overline{\mathbf h}_k^H
    +\beta_k^2\mathbf V\boldsymbol\Psi_k\mathbf V^H.
    \label{eq:full-port-second-moment}
\end{equation}
The channel correlation matrices $\{\mathbf Q_k\}_{k=1}^{K}$ are computed once before the iterations and reused throughout the algorithm.
The quadratic and linear coefficients of~\eqref{eq:profiled-quadratic} are then collected in $\mathbf W\in\mathbb C^{N_{\rm s}\times N_{\rm s}}$ and $\mathbf U\in\mathbb C^{N_{\rm s}\times K}$, respectively, as
\begin{align}
    \mathbf W&=\sum_{k=1}^{K}\nu_k|\mu_k|^2\mathbf Q_k
    +\frac{\sigma^2}{P_{\rm T}}\!\left(\sum_{k=1}^{K}\nu_k|\mu_k|^2\right)\mathbf I_{N_{\rm s}},
    \label{eq:score-matrix-W}\\
    \mathbf U&=[\nu_1\mu_1\overline{\mathbf h}_1,\ldots,
    \nu_K\mu_K\overline{\mathbf h}_K].
    \label{eq:score-matrix-U}
\end{align}
The matrices $\mathbf W$ and $\mathbf U$ are formed once per position update and shared by all candidate configurations.
For a feasible position set $\mathcal S$, let $\mathbf E_{\mathcal S}\in\mathbb R^{N_{\rm s}\times n_{\rm s}}$ be its selection matrix. The corresponding principal submatrix of $\mathbf W$ and selected rows of $\mathbf U$ are given by
\begin{align}
    \mathbf W_{\mathcal S\mathcal S}
    &=\mathbf E_{\mathcal S}^T\mathbf W\mathbf E_{\mathcal S}
    \in\mathbb C^{n_{\rm s}\times n_{\rm s}},
    \label{eq:selected-coefficient-matrices}\\
    \mathbf U_{\mathcal S}
    &=\mathbf E_{\mathcal S}^T\mathbf U
    \in\mathbb C^{n_{\rm s}\times K}.
    \label{eq:selected-linear-coefficient-matrix}
\end{align}
Using these definitions,~\eqref{eq:profiled-quadratic} can be expressed as
\begin{equation}
\begin{aligned}
    f(\mathbf E_{\mathcal S},\mathbf P)
    =\operatorname{Tr}\!\left(
    \mathbf P^H\mathbf W_{\mathcal S\mathcal S}
    \mathbf P\right)-2\operatorname{Re}\!\left\{
    \operatorname{Tr}\!\left(
    \mathbf U_{\mathcal S}^H\mathbf P
    \right)\right\}.
\end{aligned}
\label{eq:set-matrix-objective}
\end{equation}\par
\endgroup


\begingroup
\color{black}
\begin{thm}\label{prop:posterior-position-score-review}
For any feasible position set $\mathcal S$, the minimum value of the robust objective in~\eqref{eq:set-matrix-objective} is $-\Gamma(\mathcal S)$, with
\begin{equation}
    \Gamma(\mathcal S)
    \triangleq\operatorname{Tr}\!\left(
    \mathbf U_{\mathcal S}^{H}
    \mathbf W_{\mathcal S\mathcal S}^{-1}
    \mathbf U_{\mathcal S}\right).
    \label{eq:port-set-score}
\end{equation}
The minimum is uniquely attained at
\begin{equation}
    \mathbf P(\mathcal S)
    =\mathbf W_{\mathcal S\mathcal S}^{-1}\mathbf U_{\mathcal S}.
    \label{eq:selection-matrix-precoder}
\end{equation}

\end{thm}

\begin{pf}
\textit{See Appendix~\ref{app:proof-position-score}.}
\end{pf}

We refer to $\Gamma(\mathcal S)$ as the score of the feasible position set $\mathcal S$. For fixed receiver coefficients and weights, a larger $\Gamma(\mathcal S)$ corresponds to a smaller value of the robust quadratic objective in~\eqref{eq:set-matrix-objective}. The feasible candidate position sets can therefore be compared by maximizing $\Gamma(\mathcal S)$ at each iteration. However, exhaustively evaluating the scores of all feasible position sets can still be computationally prohibitive. To address this issue, we next develop a low-complexity sequential position-replacement method.

\endgroup


\subsubsection{Sequential Position Replacement via Incremental Scoring}
\red{\blu{We next optimize the MA positions sequentially, updating one position at a time while keeping the others fixed.} Let $s_i\in\mathcal S$ denote the selected position index being updated and $\mathcal T=\mathcal S\setminus\{s_i\}$ the set of retained position indices, with $|\mathcal T|=n_{\rm s}-1$. The feasible replacements for $s_i$ are given by
\begin{equation}
\mbox{\small$\displaystyle
\begin{aligned}
    \mathcal C(\mathcal T)\!=\!\bigl\{n\!\in\!\{1,\ldots,N_{\rm s}\}\!\setminus\!\mathcal T\!:\!
    \|\mathbf r_n\!-\!\mathbf r_a\|_2\ge d_{\min}, \forall a\in\mathcal T\bigr\}.
\end{aligned}
$}
\label{eq:swap-feasible-candidates}
\end{equation}
Since $s_i\in\mathcal C(\mathcal T)$, retaining the current position is always allowed. For candidate $n$, define the vectors $\mathbf u_n\triangleq[\mathbf U]_{n,:}^H\in\mathbb C^{K\times1}$ and $\mathbf w_n(\mathcal T)\triangleq[\mathbf W]_{\mathcal T,n}\in\mathbb C^{|\mathcal T|\times1}$. }
\endgroup
\begingroup\color{black}
Let $\Gamma(\mathcal T)$ denote the score associated with the retained position set $\mathcal T$, The incremental score obtained by adding candidate position $n$ to $\mathcal T$ is defined as
\begin{equation}
    \Delta_n(\mathcal T)
    \triangleq\Gamma(\mathcal T\cup\{n\})-\Gamma(\mathcal T).
    \label{eq:incremental-score-definition}
\end{equation}
Since $\Gamma(\mathcal T)$ is common to all candidates, maximizing $\Gamma(\mathcal T\cup\{n\})$ is equivalent to maximizing $\Delta_n(\mathcal T)$. We therefore select the feasible replacement with the largest incremental score:
\begin{equation}
    n^\star=\arg\max_{n\in\mathcal C(\mathcal T)}\Delta_n(\mathcal T).
    \label{eq:position-exchange-rule}
\end{equation}\par
\endgroup

\begingroup\color{black}
\begin{thm}\label{prop:incremental-scoring-review}
For $n_{\rm s}>1$, the robust incremental score of any candidate $n\in\mathcal C(\mathcal T)$ is given by
\begin{equation}
    \Delta_n(\mathcal T)
    =\frac{\|\mathbf b_n(\mathcal T)\|_2^2}{\kappa_n(\mathcal T)},
    \label{eq:port-schur-gain}
\end{equation}
where
\begin{align}
    \kappa_n(\mathcal T)
    &=W_{nn}-\mathbf w_n^H(\mathcal T)
    \mathbf W_{\mathcal T\mathcal T}^{-1}\mathbf w_n(\mathcal T)>0,
    \label{eq:port-schur-denominator}\\
    \mathbf b_n(\mathcal T)
    &=\mathbf u_n-\mathbf U_{\mathcal T}^{H}
    \mathbf W_{\mathcal T\mathcal T}^{-1}\mathbf w_n(\mathcal T).
    \label{eq:port-schur-residual}
\end{align}
For $n_{\rm s}=1$, setting $\Gamma(\varnothing)=0$ gives $\Delta_n(\varnothing)=\frac{\|\mathbf u_n\|_2^2}{W_{nn}}$.
\end{thm}
\endgroup

\begingroup\color{black}
\begin{pf}
\textit{See Appendix~\ref{app:proof-incremental-score}.}
\end{pf}
\endgroup

\gre{
For nonempty $\mathcal T$, the closed-form expression enables all candidate incremental scores to be evaluated using a shared factorization of $\mathbf W_{\mathcal T\mathcal T}$, avoiding a separate matrix inversion for each candidate, thereby reducing the computational cost of each position update.
}

\gre{
For the selected candidate set $\mathcal S_{\rm cand}=\mathcal T\cup\{n^\star\}$, the transmit precoding matrix is obtained by normalizing the closed-form solution:
\begin{equation}
    \mathbf P_{\rm cand}=\sqrt{P_{\rm T}}\,
    \frac{\mathbf P(\mathcal S_{\rm cand})}
    {\|\mathbf P(\mathcal S_{\rm cand})\|_{\rm F}}.
    \label{eq:robust-power-normalization}
\end{equation}
The candidate pair is accepted only if it increases the sum-rate lower bound; otherwise, the current position set and precoding matrix are retained.
}

\begin{algorithm}[!t]
    \color{black}
    \makeatletter
    \renewcommand{\fnum@algorithm}{\textcolor{black}{\fname@algorithm~\thealgorithm}}
    \makeatother
    \caption{\textcolor{black}{RPA-WMMSE Algorithm}}
    \label{alg:overall_ao}
    \small
    \renewcommand{\algorithmicdo}{}
    \renewcommand{\algorithmicthen}{}
    \begin{spacing}{1}
    \begin{algorithmic}[1]
        \STATE \textbf{Input:} $\{\overline{\mathbf h}_k,\beta_k,\mathbf m_k\}_{k=1}^{K}$, $\mathbf V$, $\sigma^2$, $P_{\rm T}$, $\{\mathbf r_n\}_{n=1}^{N_{\rm s}}$, $n_{\rm s}$, $d_{\min}$; feasible $(\mathcal S^{(0)},\mathbf P^{(0)})$, $I_{\max}$; tolerances $\varepsilon_{\rm stop}>0$.
        \STATE Compute $\{\mathbf Q_k\}_{k=1}^{K}$ using~\eqref{eq:full-port-second-moment}.
        \STATE Set $(\mathcal S,\mathbf P)=(\mathcal S^{(0)},\mathbf P^{(0)})$ and $R=R^{\rm lb}(\mathbf E_{\mathcal S},\mathbf P)$.
        \FOR{$s=1,\ldots,I_{\max}$}
            \STATE Let $R_{\rm old}=R$.
            \FOR{$i=1,\ldots,n_{\rm s}$}
                \STATE Update $\boldsymbol\mu,\boldsymbol\nu$ using~\eqref{Receiver}--\eqref{Weight}.
                \STATE Compute $\mathbf W,\mathbf U$ using~\eqref{eq:score-matrix-W}--\eqref{eq:score-matrix-U}.
                \STATE Let $\mathcal T=\mathcal S\setminus\{s_i\}$, compute $\Delta_n(\mathcal T)$ using~\eqref{eq:port-schur-gain}.
                \STATE Select $n^\star$ using~\eqref{eq:position-exchange-rule}.
                \STATE Form $\mathcal S_{\rm cand}$ by replacing $s_i$ with $n^\star$ in $\mathcal S$.
                \STATE Compute the precoding matrix $\mathbf P_{\rm cand}$ using~\eqref{eq:selection-matrix-precoder} and~\eqref{eq:robust-power-normalization}.
                \STATE Compute $R_{\rm cand}=R^{\rm lb}(\mathbf E_{\mathcal S_{\rm cand}},\mathbf P_{\rm cand})$.
                \IF{$R_{\rm cand}>R$}
                    \STATE $(\mathcal S,\mathbf P,R)\leftarrow(\mathcal S_{\rm cand},\mathbf P_{\rm cand},R_{\rm cand})$.
                \ENDIF
            \ENDFOR
            \IF{$|R-R_{\rm old}|<\varepsilon_{\rm stop}$}
                \STATE \textbf{Break}.
            \ENDIF
        \ENDFOR
        \STATE \textbf{Output:} $\mathcal S$ and $\mathbf P$.
    \end{algorithmic}
    \end{spacing}
\end{algorithm}

\subsection[Algorithm Summary and Complexity Analysis]{\blu{Algorithm Summary and Complexity Analysis}}\label{subsec:exchange-algorithm}

\blu{The RPA-WMMSE algorithm is summarized in \algref{alg:overall_ao}. Starting from a feasible solution, it sequentially updates the selected positions using incremental scores with fixed auxiliary variables. After precoding recovery and normalization, a feasible candidate is accepted only if it increases the sum-rate lower bound. The algorithm stops when the improvement over a complete scan falls below $\varepsilon_{\rm stop}$ or $I_{\max}$ scans have been completed. The nondecreasing and bounded sum-rate lower bound guarantees convergence in objective value.}

\blu{For dense matrix computations, constructing $\{\mathbf Q_k\}_{k=1}^{K}$ once before the iterations costs $\mathcal O(KN_{\rm s}^3)$. Each position update forms $\mathbf W$ and $\mathbf U$ at a cost of $\mathcal O(KN_{\rm s}^2)$. Factoring the shared matrix $\mathbf W_{\mathcal T\mathcal T}$ costs $\mathcal O(n_{\rm s}^3)$, while evaluating all candidate incremental scores costs $\mathcal O(N_{\rm s}n_{\rm s}^2)$ under $K\le n_{\rm s}\le N_{\rm s}$. Reusing this factorization avoids a separate matrix factorization for each candidate. With at most $I_{\max}$ scans, each updating $n_{\rm s}$ positions, the overall online complexity, including initialization, is
$\mathcal C_{\text{RPA-WMMSE}}=\mathcal O\bigl( KN_{\rm s}^3+I_{\max}n_{\rm s}KN_{\rm s}^2+I_{\max}N_{\rm s}n_{\rm s}^3\bigr)$.}

\endgroup

\section[Scalable Robust MA Transmission via Model-Driven Reinforcement Learning]{\textcolor{black}{Scalable Robust MA Transmission via Model-Driven Reinforcement Learning}}
\label{sec:rl-framework}

\begin{figure*}[t]
    \centering
    \includegraphics[width=0.9\textwidth]{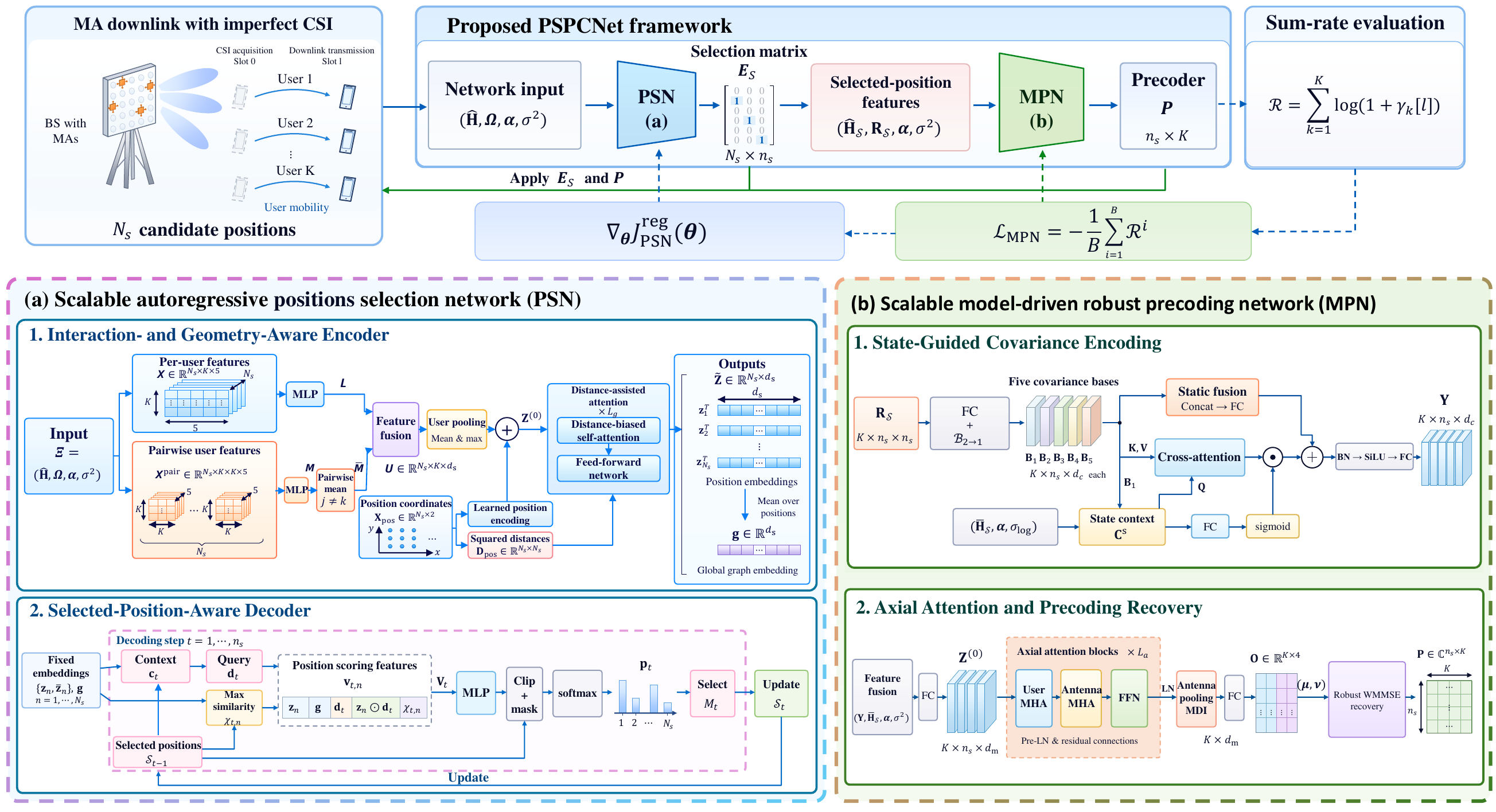}
    \caption{The overall structure of the RL framework. MLP, MHA, and FFN denote multilayer perceptron, multihead attention, and feed-forward network, respectively.}
    \label{pic:TECIZF}
    \vspace{-16pt}
\end{figure*}

\blu{In MA systems, robust joint antenna positioning and precoding constitute a nonconvex combinatorial optimization problem. \algref{alg:overall_ao} addresses this problem iteratively, with repeated candidate evaluations and precoding updates leading to high computational cost. To further improve computational efficiency, we develop PSPCNet using a reinforcement learning approach to tackle the discrete \textcolor{black}{antenna positioning} problem.} As shown in Fig.~\ref{pic:TECIZF}, the framework consists of the \textcolor{black}{position selection network (PSN) and model-driven precoding network (MPN)}. \textcolor{black}{To fully account for channel uncertainty, \blu{the PSN learns a stochastic mapping from the available CSI and channel statistics to selected position indices and is trained using the REINFORCE policy gradient method}, and the MPN combines the CSI at the selected MA positions with the corresponding channel covariances to predict key precoding parameters of the robust WMMSE structure.} \textcolor{black}{By exploiting TE processing and parameter sharing structure, PSPCNet supports robust joint \textcolor{black}{antenna positioning} and precoding across various numbers of users $K$ and MAs $n_{\rm s}$ without retraining.}

\subsection{Scalable Autoregressive Position Selection Network}
\label{subsec:pairwise-psn}

\textcolor{black}{We design an RL-based PSN that jointly exploits the available CSI and channel statistics to sequentially construct a feasible position subset. Its input is defined as}
\begin{equation}
    \boldsymbol{\Xi}
    =\big(
    \widehat{\mathbf H},\boldsymbol{\Omega},
    \boldsymbol{\alpha},\sigma^2
    \big),
    \label{eq:learning-state}
\end{equation}
where
$\widehat{\mathbf H}
=[\mathbf h_1^T[0];\ldots;\mathbf h_K^T[0]]
\in\mathbb C^{K\times N_{\rm s}}$
is the channel matrix for all candidate positions,
$\boldsymbol{\Omega}
=[\mathbf m_1^T;\ldots;\mathbf m_K^T]$
is the statistical matrix,
$\boldsymbol{\alpha}
=[\alpha_1,\ldots,\alpha_K]^T$
is the temporal correlation coefficient vector,
and $\sigma^2$ is the noise power.
The PSN parameterized by $\boldsymbol{\theta}$ autoregressively generates the position selection sequence $\mathcal S$. Let $M_t$ denote the position index selected at decoding step $t$, and $\mathcal S_t=(M_1,\ldots,M_t)$ denote the selection sequence over the first $t$ steps, where $\mathcal S_0$ is the empty sequence and $\mathcal S_{n_{\rm s}}=\mathcal S$. The corresponding stochastic selection policy is expressed as:
\begin{align}
    p_{\boldsymbol{\theta}}(\mathcal S|\boldsymbol{\Xi})
    =\prod_{t=1}^{n_{\rm s}}
    p_{\boldsymbol{\theta}}
    \big(
    M_t\,\big|\,
    \mathcal S_{t-1},\boldsymbol{\Xi}
    \big),
    \label{eq:psn-policy-factorization}
\end{align}
In fact, the mapping from $\boldsymbol{\Xi}$ to the optimal $\mathcal S^\star$ satisfies TE, meaning that when an input tensor is permuted along a dimension, the output is permuted along the corresponding dimension or remains unchanged~\cite{11049893}. Specifically, this mapping is permutation-invariant along the user dimension: simultaneously permuting the user dimensions of $\boldsymbol{\Xi}$ does not change the optimal $\mathcal S^\star$. Exploiting this property, we next design the TE-based PSN, which enables the trained model to accommodate different numbers of users $K$~\cite{11049893}.

As shown in Fig.~\ref{pic:TECIZF}(a), we implement the PSN using an autoregressive encoder--decoder architecture~\cite{huang2025rethinking,kool2019}.
\textcolor{black}{To capture multiuser interference under imperfect CSI, we design pairwise user-interaction encoding that augments per-user features with pairwise channel products and differences in channel statistics. We further introduce distance-biased spatial attention that explicitly incorporates inter-position distances into attention scores, so that feature aggregation accounts for candidate-position geometry. The decoder generates selection probabilities from the selected-position context and the maximum feature similarity to selected positions.}
Fixed-dimensional features and shared parameters enable adaptation to {different feasible numbers of MAs $n_{\rm s}$ through changes to the decoding length and feasibility mask}.

\subsubsection{Interaction- and Geometry-Aware Encoder}

At each position, the encoder combines per-user features with
aggregated pairwise interaction features to construct
the user representations~\cite{battaglia2016interaction}. Permutation-invariant pooling then produces position embeddings, which are refined through self-attention with a distance bias to incorporate relationships among positions and aperture geometry~\cite{sperber2018self}. Specifically, defining $\mathbf A_{\alpha}=\mathbf 1_{N_{\rm s}}\boldsymbol{\alpha}^T\in\mathbb R^{N_{\rm s}\times K}$ and $\sigma_{\log}=-\log_{10}(\sigma^2)$, the encoder input can be constructed as
\begin{align}
    \tX\!=\!\big[\mathfrak{R}\{\widehat{\mathbf H}^T\},\mathfrak{I}\{\widehat{\mathbf H}^T\},\boldsymbol{\Omega}^T\!\!,\sigma_{\log}\mathbf 1,\mathbf A_{\alpha}\big]_{(3)}\!\!\in\!\mathbb R^{N_{\rm s}\times K\times5}\!. \label{eq:psn-local-feature}
\end{align}
To represent interactions between users,
let $\mathbf C_n
=\widehat{\mathbf H}_{:,n}\widehat{\mathbf H}_{:,n}^H
\in\mathbb C^{K\times K}$ and
$\Delta(\mathbf a)=[a_k-a_j]_{k,j=1}^{K}
\in\mathbb R^{K\times K}$
for $\mathbf a\in\mathbb R^K$,
the pairwise feature tensor
$\tX_n^{\rm pair}\in\mathbb R^{K\times K\times5}$ at position $n$ is
\begin{equation}
\begin{aligned}
\tX_n^{\rm pair}\!\!=\!\!\big[
\mathfrak R\{\mathbf C_n\},\!\mathfrak I\{\mathbf C_n\},\!
\Delta(|\widehat{\mathbf H}_{:,n}|),\!
\Delta(\boldsymbol{\Omega}_{:,n}),\!\Delta(\boldsymbol{\alpha})
\big]_{(3)}.
\end{aligned}
\label{eq:psn-pair-port-feature}
\end{equation}
The $\mathfrak R\{\mathbf C_n\}$ and
$\mathfrak I\{\mathbf C_n\}$ jointly encode channel amplitude
and relative-phase information.
The difference matrices
$\Delta(|\widehat{\mathbf H}_{:,n}|)$,
$\Delta(\boldsymbol{\Omega}_{:,n})$, and
$\Delta(\boldsymbol{\alpha})$ capture pairwise differences
in channel amplitude, statistical amplitude, and temporal
correlation coefficient, respectively.
Stacking over positions gives 
$\tX^{\rm pair}
=[\tX_1^{\rm pair},\ldots,\tX_{N_{\rm s}}^{\rm pair}]_{(1)}
\in\mathbb R^{N_{\rm s}\times K\times K\times5}$.

\textcolor{black}{Separate two-layer MLPs with GELU activation map $\tX$ and $\tX^{\rm pair}$ to $\tL$ and $\tM$, with output feature widths $d_{\rm u}$ and $d_{\rm p}$, respectively.
Each MLP operates along the feature dimension, with its parameters shared across positions and user indices.}
\textcolor{black}{Let $\tM'$ denote the tensor obtained from $\tM$ by removing the diagonal entries corresponding to self-interactions. The interactions with other users are then averaged and fused with the per-user features as}
\begin{align}
    \tU\!
    =\!\operatorname{LN}\!\left(
    \operatorname{GELU}\!\left(
    \mathrm{FC}\!\left(
    [\tL,\operatorname{Mean}_{3}(\tM')]_{3}
    \right)\right)\right)
    \!\in\!\mathbb R^{N_{\rm s}\times K\times d_{\rm s}}\!,
    \label{eq:pair-message}
\end{align}
where \textcolor{black}{$\mathrm{FC}$ denotes a fully connected layer,} $\operatorname{LN}$ denotes layer normalization\textcolor{black}{,} and $d_{\rm s}$ is the main feature width of the PSN. The shared feature processing and mean aggregation preserve permutation equivariance along the user dimension~\cite{11049893}.

Mean and max pooling over users produce position representations invariant to user permutations. To exploit the spatial channel correlation, we further add a learnable
positional encoding to
incorporate aperture geometry, which yields
\begin{equation}
\begin{aligned}
\mathbf Z^{(0)}
&=\mathrm{FC}\!\left(
[\operatorname{Mean}_2(\tU),
\operatorname{Max}_2(\tU)]_2
\right)\\
&\quad+\mathrm{FC}\!\left(
\operatorname{Tanh}\!\left(
\mathrm{FC}(\mathbf X_{\rm pos})
\right)\right)
\in\mathbb R^{N_{\rm s}\times d_{\rm s}},
\end{aligned}
\label{eq:user-invariant-pooling}
\end{equation}
where $\mathbf X_{\rm pos}\in\mathbb R^{N_{\rm s}\times2}$
contains the two-dimensional candidate-position coordinates, with each
coordinate dimension linearly mapped to $[-1,1]$.

The encoder then employs $L_{\rm g}$ distance-assisted attention blocks to aggregate feature information among candidate positions~\cite{sperber2018self}. Let $\mathbf D_{\rm pos}\in\mathbb R^{N_{\rm s}\times N_{\rm s}}$ denote the matrix of squared distances between candidate positions, where $[\mathbf D_{\rm pos}]_{i,j}=\|[\mathbf X_{\rm pos}]_{i,:}-[\mathbf X_{\rm pos}]_{j,:}\|_2^2$.
The $l$-th attention block takes $\mathbf Z^{(l-1)}$ as input and generates $\mathbf Q^{(l)},\mathbf K^{(l)},\mathbf V^{(l)}\in\mathbb R^{N_{\rm s}\times d_{\rm s}}$ through layer normalization and three linear projections. The corresponding attention matrix is
\begin{align}
    \mathbf A^{(l)}
    &=\operatorname{softmax}\!\left(
    \frac{\mathbf Q^{(l)}{\mathbf K^{(l)}}^T}{\sqrt{d_{\rm s}}}
    -\operatorname{softplus}(\rho_l)\mathbf D_{\rm pos}
    \right),
    \label{eq:radial-graph-attention}
\end{align}
where $\rho_l$ is a learnable parameter. The distance term reduces the attention scores between more distant positions, so that feature aggregation accounts for both node feature correlations and aperture geometry. The block then updates the position embeddings
$\mathbf Z^{(l-1)}$ to $\mathbf Z^{(l)}$
through attention aggregation and a feed-forward network
with GELU activation, using pre-normalization and residual
connections~\cite{xiong2020layernorm}. After $L_{\rm g}$ attention blocks, the encoder outputs the position node embeddings and the global graph embedding:
\begin{equation}
    \begin{aligned}
        \widetilde{\mathbf Z}
        =\operatorname{LN}\!\left(\mathbf Z^{(L_{\rm g})}\right)
        \in\mathbb R^{N_{\rm s}\times d_{\rm s}},\ \ 
        \mathbf g
        =\operatorname{Mean}_{1}(\widetilde{\mathbf Z})
        \in\mathbb R^{d_{\rm s}}.
    \end{aligned}
    \label{eq:psn-encoder-output}
\end{equation}

\subsubsection{Selected-Position-Aware Decoder}

Using the encoder outputs $\widetilde{\mathbf Z}$ and
$\mathbf g$, the decoder sequentially selects positions conditioned on the previously selected positions.
We define
\begin{equation}
\mathbf z_n=[\widetilde{\mathbf Z}]_{n,:}^T\in\mathbb R^{d_{\rm s}},
\qquad
\overline{\mathbf z}_n=\frac{\mathbf z_n}{\max\{\|\mathbf z_n\|_2,\varepsilon\}}\in\mathbb R^{d_{\rm s}},
\label{eq:psn-node-normalization}
\end{equation}
where $\varepsilon$ is a small positive constant that ensures
numerical stability. The $\mathbf z_n$,
$\overline{\mathbf z}_n$, and $\mathbf g$ are computed once
and reused across all $n_{\rm s}$ decoding steps~\cite{huang2025rethinking}.

At step $t$, the selected positions are summarized by their mean embedding $\mathbf c_t$, while each candidate is characterized by its maximum feature similarity $\chi_{t,n}$ to the selected positions. For $t>1$, these quantities are given by
\begin{align}
\mathbf c_t
&=\frac{1}{t-1}\sum_{r=1}^{t-1}\mathbf z_{M_r},
\label{eq:psn-selected-context}\\
\chi_{t,n}
&=\max_{1\le r<t}\overline{\mathbf z}_n^T
\overline{\mathbf z}_{M_r}.
\label{eq:psn-max-similarity}
\end{align}
At $t=1$, we set $\mathbf c_1=\mathbf c_{\varnothing}$ and $\chi_{1,n}=0$, where $\mathbf c_{\varnothing}\in\mathbb R^{d_{\rm s}}$ is learnable.

To incorporate both global information and the current selection
context into position scoring, we fuse $\mathbf g$ and $\mathbf c_t$
to form the decoding query $\mathbf d_t
=\operatorname{LN}\!\left(\mathrm{FC}\!\left(
\operatorname{GELU}\!\left(
\mathrm{FC}([\mathbf g,\mathbf c_t]_1)
\right)\right)\right)
\in\mathbb R^{d_{\rm s}}.$
The scoring feature
for position $n$ is
\begin{equation}
\mathbf v_{t,n}=
[\mathbf z_n,\mathbf g,\mathbf d_t,
\mathbf z_n\odot\mathbf d_t,\chi_{t,n}]_1 \in\mathbb R^{4d_{\rm s}+1},
\label{eq:psn-score-features}
\end{equation}
where $\mathbf z_n\odot\mathbf d_t$ captures the match between the candidate embedding and the current query.
Define
$\mathbf V_t=[\mathbf v_{t,1},\ldots,\mathbf v_{t,N_{\rm s}}]^T
\in\mathbb R^{N_{\rm s}\times(4d_{\rm s}+1)}$
and the score vector is obtained as
$\mathbf s_t=\mathrm{FC}(\operatorname{GELU}
(\mathrm{FC}(\mathbf V_t)))\in\mathbb R^{N_{\rm s}}$.
\textcolor{black}{The conditional selection probabilities are then obtained as}
\begin{equation}
\mathbf p_t=\operatorname{softmax}\!\left(C\tanh(\mathbf s_t)+\mathbf m_t\right),
\label{eq:psn-step-probability}
\end{equation}
where $C>0$ is the score clipping coefficient and $[\mathbf m_t]_n=-\infty$ for selected positions or candidates violating the minimum-spacing constraint, and $0$ otherwise~\cite{kool2019}.
The decoder samples $M_t$ from $\mathbf p_t$ during training and chooses the highest-probability position during inference~\cite{kool2019}.
The context $\mathbf c_{t+1}$,
similarity features $\chi_{t+1,n}$, and mask
$\mathbf m_{t+1}$ are updated after each selection
and the same decoding module then selects the next position.
This procedure continues until the $n_{\rm s}$ positions are obtained.
Since the feature dimensions are independent of $n_{\rm s}$ and the decoder parameters are
shared across steps, the same trained decoder supports different feasible numbers of MAs by adjusting the decoding length.

\subsection[Scalable Model-Driven Robust Precoding Network]{\textcolor{black}{Scalable Model-Driven Robust Precoding Network}}
\label{subsec:state-conditioned-mpn}

\textcolor{black}{Given a feasible position selection sequence $\mathcal S$ generated by the PSN, we consider the corresponding robust precoding subproblem in~\eqref{eq:problem wmmse 2}. Existing robust precoding networks for UPAs do not account for position-dependent changes in the effective channel and its covariance in MA systems~\cite{shi2021dlrobust}. To this end, we design a TE model-driven precoding network that fuses the selected-position channel information and covariance through CSI-conditioned equivariant encoding. It predicts the key parameters $\boldsymbol{\mu}$ and $\boldsymbol{\nu}$ for closed-form robust WMMSE recovery using~\eqref{eq:selection-matrix-precoder}~\cite{shi2023robustwmmse}. The corresponding parameter mapping is expressed as}
\begin{equation}
(\boldsymbol{\mu}^\star,\boldsymbol{\nu}^\star)
=G\!\left(\widehat{\mathbf H}_{\mathcal S},
\{\mathbf R_{k,\mathcal S}\}_{k=1}^K,
\boldsymbol{\alpha},\sigma^2\right),
\label{eq:mpn-parameter-map}
\end{equation}
where $\widehat{\mathbf H}_{\mathcal S}=\widehat{\mathbf H}\mathbf E_{\mathcal S}\in\mathbb C^{K\times n_{\rm s}}$ is the selected-position channel, and $\mathbf R_{k,\mathcal S}=\mathbf E_{\mathcal S}^T\mathbf R_{k}\mathbf E_{\mathcal S}$ is its channel covariance. \textcolor{black}{The mapping $G$ is equivariant to user permutations and invariant to a common permutation of the selected positions in $\widehat{\mathbf H}_{\mathcal S}$ and in both dimensions of each $\mathbf R_{k,\mathcal S}$~\cite{11049893}.} We exploit these properties in the TE MPN to enable parameter reuse across different numbers of users $K$ and MAs $n_{\rm s}$~\cite{11049893}. 

The MPN architecture is illustrated in Fig.~\ref{pic:TECIZF}(b).
\textcolor{black}{To exploit position-dependent channel statistics for robust precoding, we design a CSI-conditioned equivariant covariance encoder. The encoder first forms a static representation from features describing channel variances, inter-antenna covariance, and global statistics. It then uses CSI-conditioned attention to reweight these features and adds the attention output to the static representation through a learned gate. The resulting covariance features are combined with the channel features and processed by axial attention along the user and antenna dimensions.}

\subsubsection{State-Guided Covariance Encoding}

Define $\tR_{\mathcal S}\in\mathbb R^{K\times n_{\rm s}\times n_{\rm s}}$ with $[\tR_{\mathcal S}]_{k,:,:}=\mathbf R_{k,\mathcal S}$. A fully connected layer first embeds each covariance element into a $d_{\rm c}$-dimensional feature space. The TE second-to-first-order feature construction in~\cite{11049893} is applied along the two $n_{\rm s}$-dimension to obtain
\begin{align}
    \tB
    =\mathcal B_{2\rightarrow1}\!\left(
    \operatorname{FC}\!\left(\tR_{\mathcal S}\right)
    \right)
    \in\mathbb R^{K\times n_{\rm s}\times5\times d_{\rm c}}.
    \label{eq:mpn-five-basis-tensor}
\end{align}
The operator $\mathcal B_{2\rightarrow1}(\cdot)$ extracts
five types of features in order: diagonal features,
the diagonal mean, row means, column means, and the global mean.
These features provide per-antenna features
and global statistical information. Let $\tB_b=[\tB]_{:,:,b,:}
\in\mathbb R^{K\times n_{\rm s}\times d_{\rm c}}$, $b=1,\ldots,5$,
denote the five types of features. We fuse these features
through a static branch and an attention branch.
The static branch concatenates them along the feature
dimension to obtain
$\tB^{\rm st}
=\operatorname{FC}([\tB_1,\ldots,\tB_5]_3)$. The attention branch complements the static representation by adaptively weighting the five types of features according to the available CSI. Define the channel mean
as $\overline{\mathbf H}_{\mathcal S}
=\operatorname{diag}(\boldsymbol{\alpha})
\widehat{\mathbf H}_{\mathcal S}$.
We combine the diagonal features $\tB_1$ with the
available CSI to construct the query context
$\tC^{\rm s}
=$$\operatorname{SiLU}\!\left(\operatorname{FC}\!\left(
\bigl[
[\mathfrak R\{\overline{\mathbf H}_{\mathcal S}\},
\mathfrak I\{\overline{\mathbf H}_{\mathcal S}\},
\boldsymbol{\alpha}\mathbf1_{n_{\rm s}}^T,
\sigma_{\log}\mathbf1]_{(3)},\tB_1
\bigr]_3
\right)\right)$$
\in\mathbb R^{K\times n_{\rm s}\times d_{\rm c}}$.
The query is obtained as
$\tQ^{\rm att}=\operatorname{FC}(\tC^{\rm s})$,
while the key $\tK^{\rm att}=\operatorname{FC}(\tB)$
and value $\tV^{\rm att}=\operatorname{FC}(\tB)$
are generated from $\tB$.
The attention weights
$\tA^{\rm att}\in\mathbb R^{K\times n_{\rm s}\times5}$
are then computed by matching the query $\tQ^{\rm att}$
with the key $\tK^{\rm att}$~\cite{zhuang2026aliasfree},
and the values are aggregated with these weights to obtain
$\tB^{\rm att}
=\operatorname{FC}(\tA^{\rm att}\tV^{\rm att})$.
Finally, $\tC^{\rm s}$ is used to gate the attention
output and fuse it with the static branch, yielding the
covariance encoding features
$\tY\in\mathbb R^{K\times n_{\rm s}\times d_{\rm c}}$:
\begin{equation}
\tY\!=\!\operatorname{FC}\!\left(
\operatorname{SiLU}\!\left(
\operatorname{BN}\!\left(
\tB^{\rm st}
\!\!+\!\operatorname{sigmoid}(\operatorname{FC}(\tC^{\rm s}))
\!\odot\!\tB^{\rm att}
\right)\!\right)\!\right)\!,
\label{eq:mpn-covariance-encoding}
\end{equation}
where $\operatorname{BN}$ denotes batch normalization. The attention mechanism adjusts the relative contributions
of the five features according to the available
CSI, allowing the statistical representation to adapt to
the channel conditions. The resulting per-antenna
features $\tY$ remain aligned with the channel
features, and the encoding preserves permutation equivariance
along the user and selected-position dimensions.

\subsubsection{Axial Attention and Precoding Recovery}

We next concatenate the obtained covariance features
$\tY$ with the channel mean
$\overline{\mathbf H}_{\mathcal S}$, noise power $\sigma^2$,
and temporal correlation coefficients $\boldsymbol{\alpha}$
to form the fused feature tensor:
\begin{equation}
\tF\!=\!\!\big[
\mathfrak R\{\overline{\mathbf H}_{\mathcal S}\!\},
\mathfrak I\{\overline{\mathbf H}_{\mathcal S}\!\},
\tY,\sigma^2\mathbf1,
\boldsymbol{\alpha}\mathbf1_{n_{\rm s}}^T\big]_3\!\!\in\!\mathbb R^{K\times n_{\rm s}\times(d_{\rm c}+4)}\!.
\label{eq:mpn-fused-features}
\end{equation}
A fully connected layer then maps the feature dimension
of $\tF$ from $d_{\rm c}+4$ to $d_{\rm m}$, yielding
$\tZ^{(0)}=\operatorname{FC}(\tF)
\in\mathbb R^{K\times n_{\rm s}\times d_{\rm m}}$.
\textcolor{black}{The features are processed by $L_{\rm a}$ axial attention blocks~\cite{ho2019axial,zhuang2026aliasfree}, each applying multihead self-attention sequentially along the user and antenna dimensions, followed by a two-layer FFN with SiLU activation. All sublayers use pre-normalization and residual connections. This processing preserves TE while avoiding a full $(Kn_{\rm s})\times(Kn_{\rm s})$ attention matrix.}

\textcolor{black}{After final layer normalization, the encoded features $\tZ^{(L_{\rm a})}$ are pooled along the antenna dimension by an attention-based multidimensional invariant (MDI) module~\cite{11049893}. A fully connected layer then generates the key precoding parameters:}
\begin{equation}
\begin{aligned}
\mathbf O&=\operatorname{FC}\!\left(
\operatorname{MDI}_2(\tZ^{(L_{\rm a})})\right)
\in\mathbb R^{K\times4},\\
\boldsymbol{\mu}&=[\mathbf O]_{:,1}+j[\mathbf O]_{:,2},
\quad
\boldsymbol{\nu}
=[\mathbf O]_{:,3}+j[\mathbf O]_{:,4},
\end{aligned}
\label{eq:mpn-parameter-output}
\end{equation}
where $\operatorname{MDI}_2(\cdot)$ denotes pooling along the second dimension. {\color{black}Using the predicted parameters $\boldsymbol\mu$ and $\boldsymbol\nu$ and the selected position set $\mathcal S$, we can construct $\mathbf W_{\mathcal S\mathcal S}$ and $\mathbf U_{\mathcal S}$ and recover the precoding matrix using~\eqref{eq:selection-matrix-precoder}. The resulting precoding matrix is then normalized using~\eqref{eq:robust-power-normalization}.}

\subsection{Staged Label-Free Training}
\label{subsec:staged-training}

We first pretrain MPN independently and then jointly optimize PSN and MPN. Both stages directly optimize the sum rate without requiring optimal solution labels.

\subsubsection{MPN Pretraining with Random Positions}

To enable MPN to learn robust precoding across different position combinations and avoid overfitting to a fixed selection pattern, we pretrain it using randomly selected positions.
For each sample, the sum rate $\mathcal R=\sum_{k=1}^{K}\log(1+\gamma_k[l])$ is evaluated using the precoder recovered via~\eqref{eq:selection-matrix-precoder} and the transmission-channel realization $\mathbf H[l]$, where $\gamma_k[l]$ is computed using~\eqref{eq:instantaneous-sinr}. For a batch of $B$ samples, MPN minimizes~\cite{11049893}
\begin{equation}
    \mathcal L_{\rm MPN}
    =-\frac{1}{B}\sum_{i=1}^{B}\mathcal R^i,
    \label{eq:mpn-rate-loss}
\end{equation}
where $i$ indexes the training samples.
\textcolor{black}{The channel realization $\mathbf H[l]$ is used only to evaluate the training loss and reward, and is not a network input.}

\subsubsection{{Joint Optimization of PSN and MPN}}

{In the second stage, PSN and the pretrained MPN are jointly trained using separate optimizers.}
Given the input $\boldsymbol{\Xi}$ defined in~\eqref{eq:learning-state},
PSN samples a position selection sequence $\mathcal S$ and MPN
generates the corresponding robust precoding matrix $\mathbf P$. PSN is
trained to maximize the expected sum rate, i.e.,
\begin{equation}
    \max_{\boldsymbol{\theta}}\
    J_{\rm PSN}(\boldsymbol{\theta})
    =\mathbb E_{\boldsymbol{\Xi},\mathbf H[l]}\!\left[
    \mathbb E_{\mathcal S\sim p_{\boldsymbol{\theta}}(\cdot\mid\boldsymbol{\Xi})}
    \!\left[\mathcal R(\mathcal S)\right]\right].
\label{eq:psn-expected-objective}
\end{equation}
Since discrete position sampling prevents direct backpropagation of the sum-rate objective, we adopt the REINFORCE policy gradient method~\cite{williams1992reinforce,kool2019}. To reduce gradient variance, a frozen reference copy of the PSN–MPN model provides a greedy rollout baseline $b$~\cite{kool2019}, evaluated using the same input and transmission-channel realization.
To prevent premature concentration of the position selection probabilities, we introduce conditional policy entropy~\cite{mnih2016asynchronous}:
\begin{equation}
\begin{aligned}
    \mathcal H
    ={}&-\sum_{t=1}^{n_{\rm s}}\sum_{n\in\mathcal A_t}
    p_{\boldsymbol{\theta}}\!\left(
    M_t=n\mid\mathcal S_{t-1},\boldsymbol{\Xi}
    \right)\\
    &\quad\times\log p_{\boldsymbol{\theta}}\!\left(
    M_t=n\mid\mathcal S_{t-1},\boldsymbol{\Xi}
    \right),
\end{aligned}
\label{eq:psn-conditional-entropy}
\end{equation}
where $\mathcal A_t=\{n:[\mathbf m_t]_n=0\}$ is the feasible candidate set at step $t$.
For a batch of $B$ training samples, the gradient of the entropy-regularized PSN objective $J_{\rm PSN}^{\rm reg}$ is estimated using the greedy rollout baseline as follows:
\begin{equation}
\begin{aligned}
    &\nabla_{\boldsymbol{\theta}}J_{\rm PSN}^{\rm reg}(\boldsymbol{\theta})
    =\frac{1}{B}\sum_{i=1}^{B}\Bigl[
    (\mathcal R^i-b^i)\\
    &\qquad\sum_{t=1}^{n_{\rm s}}\nabla_{\boldsymbol{\theta}}
    \log p_{\boldsymbol{\theta}}\!\left(
    M_t^i\mid\mathcal S_{t-1}^i,\boldsymbol{\Xi}^i
    \right)+\eta_{\rm H}
    \nabla_{\boldsymbol{\theta}}\mathcal H^i\Bigr],
\end{aligned}
\label{eq:joint-psn-loss}
\end{equation}
The entropy coefficient $\eta_{\rm H}$ gradually decreases during training, shifting the training focus from exploring position combinations to optimizing the sum rate.

{In each training batch, PSN is updated using the policy gradient in~\eqref{eq:joint-psn-loss}, and MPN minimizes the sum-rate loss in~\eqref{eq:mpn-rate-loss}. During the first epoch, the batch-mean reward is used as the baseline, and both networks are copied to the reference model. In subsequent epochs, both PSN and MPN parameters are copied to the reference model only when the current model achieves a higher average sum rate and passes a one-sided paired $t$-test ($p<\epsilon$)~\cite{kool2019}. The reference model remains frozen between these updates.}

\subsection{Computational Complexity}
\label{subsec:advantages-complexity}

For fixed network depths, the computational complexity of PSPCNet mainly comes from PSN and MPN inference, channel covariance construction, and closed-form precoding recovery. Under $1\le K\le n_{\rm s}\le N_{\rm s}$, these operations yield an overall complexity of
$\mathcal C_{\rm RL}=\mathcal O\bigl(N_{\rm s}K^2d_{\rm s}^2+n_{\rm s}N_{\rm s}d_{\rm s}(N_{\rm s}+d_{\rm s})+Kn_{\rm s}d_{\rm m}(n_{\rm s}+d_{\rm m})+KN_{\rm s}n_{\rm s}^2\bigr),$
where $d_{\rm s}$ and $d_{\rm m}$ denote the main feature widths of PSN and MPN, respectively, with the remaining layer widths proportional to the corresponding main widths. The first two terms correspond to PSN, the third to MPN, and the last to covariance construction and precoding recovery.
\blu{Compared with RPA-WMMSE in Section~\ref{subsec:exchange-algorithm}, PSPCNet performs one sequential position-selection pass followed by one closed-form precoding recovery, eliminating repeated optimization scans.}
 Moreover, parameter sharing keeps the number of trainable parameters independent of $K$ and $n_{\rm s}$ for fixed network widths and depths.

\begin{table}[t]
    \centering
    \caption{Network and Training Parameters}
    \label{tab:network-training-parameters}
    \footnotesize
    \setlength{\tabcolsep}{3pt}
    \renewcommand{\arraystretch}{1.15}
    \begin{tabular}{
        @{}
        >{\raggedright\arraybackslash}p{0.70\columnwidth}
        >{\raggedright\arraybackslash}p{0.25\columnwidth}
        @{}
    }
        \toprule
        \textbf{Parameter} & \textbf{Value} \\
        \midrule
        \multicolumn{2}{@{}l}{\textbf{PSN}} \\
        Feature widths $(d_{\rm s},d_{\rm u},d_{\rm p})$
            & $(128,64,32)$ \\
        Distance-assisted attention blocks $L_{\rm g}$
            & $2$ \\
        Query, scoring, and FFN hidden width
            & $2d_{\rm s}$ \\
        Score clipping coefficient $C$
            & $10$ \\
        \midrule
        \multicolumn{2}{@{}l}{\textbf{MPN}} \\
        Feature widths $(d_{\rm m},d_{\rm c})$
            & $(32,4)$ \\
        Axial attention blocks $L_{\rm a}$
            & $2$ \\
        FFN hidden width
            & $2d_{\rm m}$ \\
        \midrule
        \multicolumn{2}{@{}l}{\textbf{Training}} \\
        Optimizer
            & Adam \\
        Baseline significance threshold $\epsilon$
            & $0.05$ \\
        \bottomrule
    \end{tabular}
\end{table}
\section{Numerical Results}\label{Section 6}
In this section, we evaluate the performance of the proposed
methods via Monte Carlo simulations. To generate channels
compliant with 3GPP standards, we utilize the widely adopted
QuaDRiGa channel simulator~\cite{3gpp_tr_38901_v190,6758357}. The BS is equipped with MAs whose positions are selected from $N_{\rm s}=100$
candidate positions arranged on a $10\times10$ uniform grid
over a $\lambda\times\lambda$ aperture, with a spacing
of $\lambda/9$ along each axis. \textcolor{black}{The minimum spacing is set to $d_{\min}=0.3\lambda$.} Unless otherwise specified,
$n_{\rm s}=4$ MAs serve $K=3$ single-antenna users. The carrier frequency is $3.5$ GHz, and users are randomly distributed in a $120^\circ$ sector of radius $500$ m in front of the BS. We adopt the 3GPP TR 38.901 urban macrocell (UMa)
channel models~\cite{3gpp_tr_38901_v190} for both non-line-of-sight (NLOS) and line-of-sight (LOS) propagation.
The NLOS tests use a user
speed of $3$ km/h and a $10$ ms interval between CSI
acquisition and downlink transmission, while the
corresponding settings for the LOS tests are $5$ km/h
and $15$ ms. To separate the effects of large-scale fading and more clearly compare different methods, shadow fading and path loss are excluded from the simulations. The channel is normalized so that its squared Frobenius norm is $KN_{\rm s}$. The statistical amplitudes $\{\mathbf m_k\}_{k=1}^{K}$
and temporal correlation coefficients $\{\alpha_k\}_{k=1}^{K}$ are estimated
from historical channel snapshots. The signal-to-noise ratio (SNR) is defined as $\mathrm{SNR}=P_{\rm T}/\sigma^2$, and $P_{\rm T}$ is set to $1$.

For each scenario, $50{,}000$ channel samples are generated,
with $45{,}000$ used for training and $5{,}000$ for testing.
Both PSN and MPN are trained across all SNR levels in
$\{0,5,\ldots,40\}$ dB.
The network architecture and fixed training parameters
are summarized in Table~\ref{tab:network-training-parameters}.
MPN is first pretrained using random position subsets for
$400$ epochs with a batch size of $800$ and an initial
learning rate of $3\times10^{-4}$, which is reduced to
$3\times10^{-5}$ after epoch $200$.
{PSN and the pretrained MPN are then jointly trained for
$200$ epochs, using a batch size of $200$ and separate
optimizers with a learning rate of $10^{-4}$ for each network.}
The entropy regularization coefficient $\eta_{\rm H}$
decreases linearly from $0.05$ to $0.005$ over the first
$120$ of these $200$ epochs and then remains unchanged.


\begin{figure*}[!t]
    \centering
\begin{minipage}[t]{0.24\textwidth}
        \vspace{0pt}
        \centering
        \includegraphics[width=\linewidth]{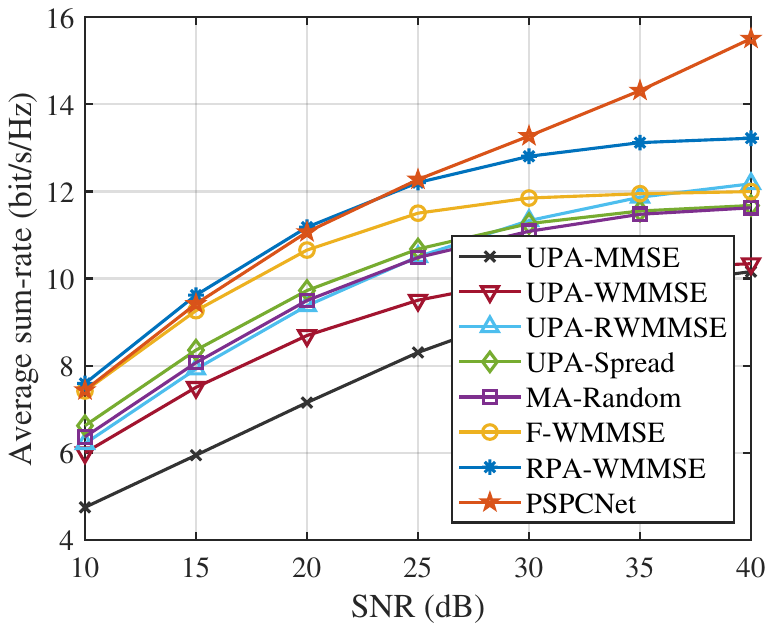}
        \caption{Average sum rate versus SNR, $n_{\rm s}=4$, $K=3$, NLOS.}
        \label{fig:sr_snr_nlos_1010}
    \end{minipage}%
    \hfill
\begin{minipage}[t]{0.24\textwidth}
        \vspace{0pt}
        \centering
        \includegraphics[width=\linewidth]{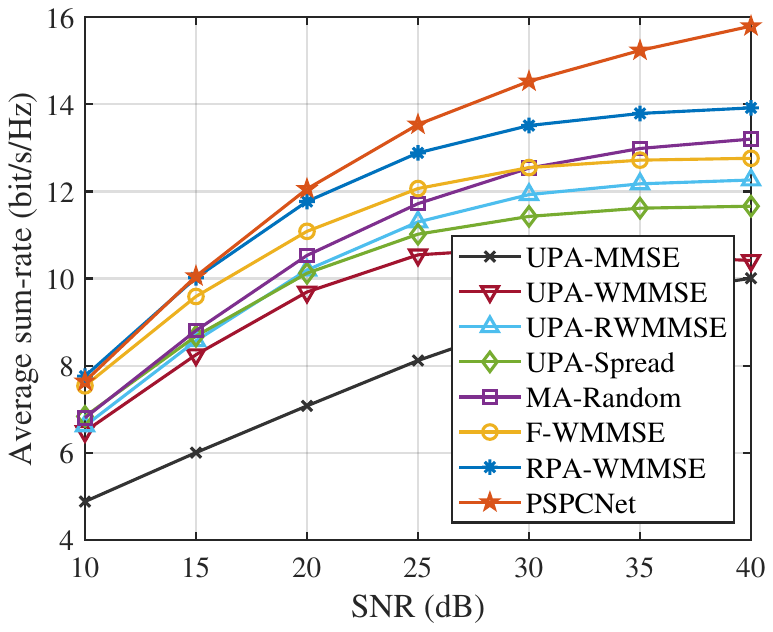}
        \caption{Average sum rate versus SNR, $n_{\rm s}=4$, $K=3$, LOS.}
        \label{fig:sr_snr_los_1010}
    \end{minipage}%
    \hfill
\begin{minipage}[t]{0.24\textwidth}
        \vspace{0pt}
        \centering
        \includegraphics[width=\linewidth]{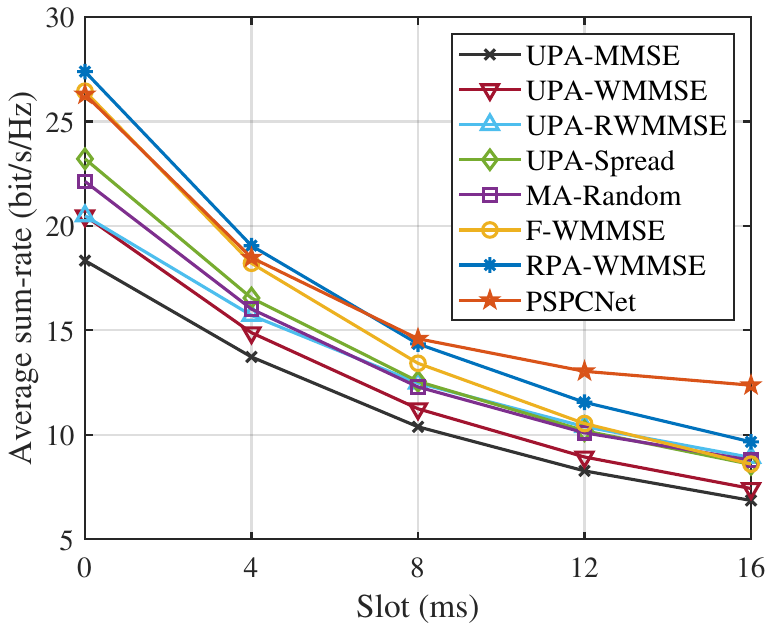}
        \caption{Average sum rate versus CSI delay, $n_{\rm s}=4$, $K=3$, NLOS, $\mathrm{SNR}=30$~dB.}
        \label{fig:sr_delay_nlos_1010}
    \end{minipage}%
    \hfill
\begin{minipage}[t]{0.24\textwidth}
        \vspace{0pt}
        \centering
        \includegraphics[width=\linewidth]{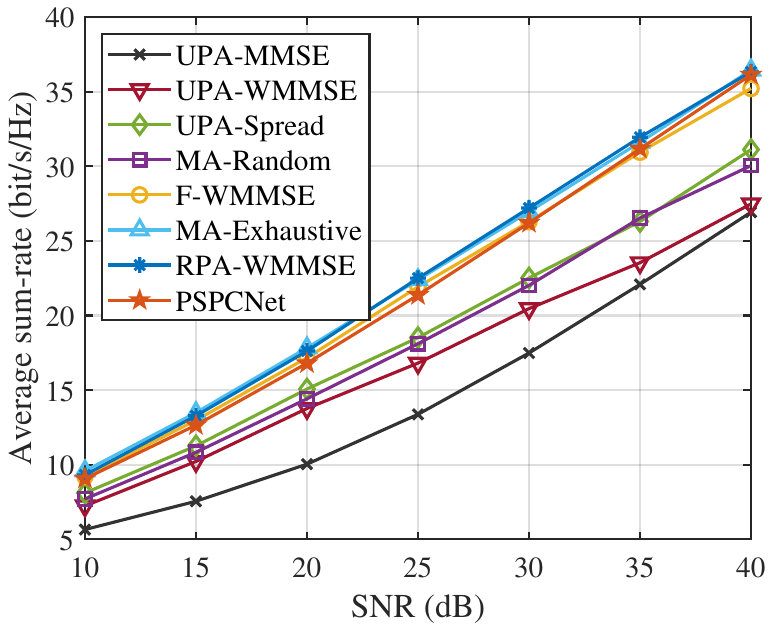}
        \caption{Average sum rate versus SNR, $n_{\rm s}=4$, $K=3$, NLOS, perfect CSI.}
        \label{fig:sr_snr_nlos_1010_perfect_csi}
    \end{minipage}
    \vspace{-12pt}
\end{figure*}

The following schemes are compared in this section:
\begin{itemize}
    \item `\textbf{UPA-RWMMSE}', `\textbf{UPA-WMMSE}', and `\textbf{UPA-MMSE}': Robust WMMSE, non-robust WMMSE, and minimum mean square error (MMSE) precoding for a conventional UPA with half-wavelength antenna spacing, respectively.
    \item `\textbf{UPA-Spread}': Robust WMMSE precoding with \textcolor{black}{a fixed antenna layout designed to maximize antenna separation within the MA moving region. The antennas occupy the four corners for $n_{\rm s}=4$ and form a $2\times3$ layout spanning the region for $n_{\rm s}=6$.}
    \item `\textbf{MA-Random}': Robust WMMSE precoding with $n_{\rm s}$ randomly selected distinct positions.
    \item `\textbf{MA-Exhaustive}': Exhaustive position search under perfect CSI, with MMSE initialization and $60$ WMMSE iterations for each combination. The highest sum rate is selected.
    \item \textcolor{black}{`\textbf{F-WMMSE}': The flexible WMMSE method in~\cite{yang2025flexiblewmmse} for joint \textcolor{black}{antenna positioning} and precoding without robust design. The iterations stop when the absolute change in the sum rate between iterations falls below $10^{-3}$ bit/s/Hz or the maximum of $100$ iterations is reached.}
    \item \textcolor{black}{`\textbf{RPA-WMMSE}': The proposed robust joint \textcolor{black}{antenna positioning} and precoding method in Section~\ref{sec:ao-robust}, with $I_{\max}=100$ and stopping tolerance $\varepsilon_{\rm stop}=10^{-3}$ bit/s/Hz.}
    \item `\textbf{PSPCNet}': The proposed PSN--MPN neural network method for robust joint \textcolor{black}{antenna positioning} and precoding in Section~\ref{sec:rl-framework}.
\end{itemize}

Figs.~\ref{fig:sr_snr_nlos_1010} and
\ref{fig:sr_snr_los_1010} compare the average sum-rate
performance of the considered schemes versus SNR in
the NLOS and LOS scenarios, respectively.
As can be observed, both RPA-WMMSE and PSPCNet
outperform the other benchmarks across the considered
SNR range, with their gains over non-robust F-WMMSE
generally becoming larger as SNR increases.
The effectiveness of \textcolor{black}{antenna positioning} is also
evident from the gains of the proposed methods over MA-Random, UPA-Spread,
and UPA-RWMMSE.
Moreover, PSPCNet achieves performance comparable to
RPA-WMMSE at low SNRs and provides further gains
at high SNRs, demonstrating the performance benefits
of the proposed network.
Notably, UPA-Spread performs worse than UPA-RWMMSE
at high SNRs, suggesting that increasing antenna spacing
alone does not necessarily improve performance and
that \textcolor{black}{antenna positioning} should also account for the
propagation environment.
Fig.~\ref{fig:sr_delay_nlos_1010} compares the average
sum rates of the considered schemes versus CSI delay.
{\leavevmode\color{black}RPA-WMMSE consistently achieves higher sum rates than
F-WMMSE. PSPCNet performs similarly to F-WMMSE
at small delays, but its advantage becomes more pronounced
as the delay increases.}
{\leavevmode\color{black}These results demonstrate the effectiveness of the
proposed robust joint \textcolor{black}{antenna positioning} and precoding
methods in mitigating the performance loss caused
by imperfect CSI.}

{\leavevmode\color{black}Fig.~\ref{fig:sr_snr_nlos_1010_perfect_csi} compares the average sum rates of the considered schemes under perfect CSI. Both RPA-WMMSE and PSPCNet achieve sum rates close to those of MA-Exhaustive and significantly outperform the random-position and fixed-array schemes. RPA-WMMSE also achieves higher sum rates than F-WMMSE, while PSPCNet performs comparably to F-WMMSE at low SNRs and slightly better at high SNRs. The slight advantage of RPA-WMMSE over MA-Exhaustive at some SNRs may arise because its precoding updates during position search reach better local solutions than those obtained from MMSE initialization.}


\begin{figure*}[!t]
    \centering
\begin{minipage}[t]{0.24\textwidth}
        \vspace{0pt}
        \centering
        \includegraphics[width=\linewidth]{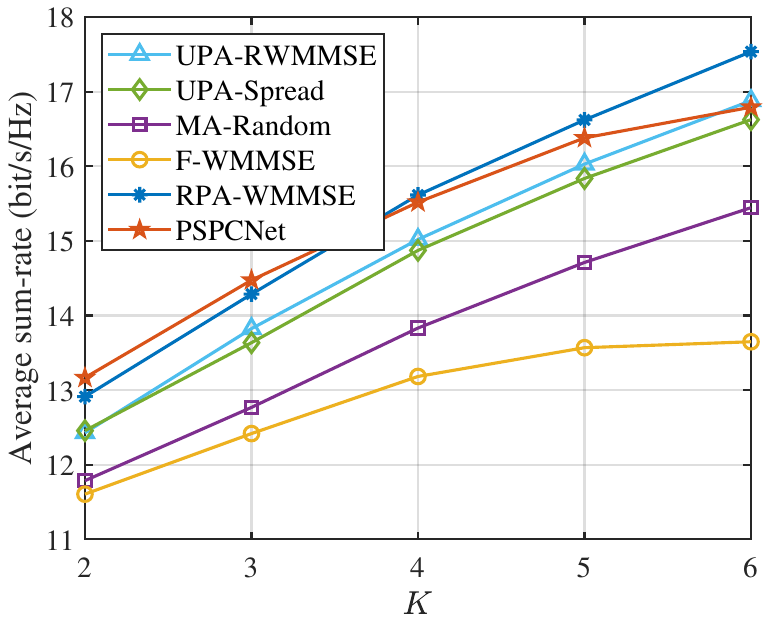}
        \caption{Average sum rate versus $K$, $n_{\rm s}=6$, NLOS, $\mathrm{SNR}=40$~dB, PSPCNet trained at $K=4$.}
        \label{fig:sr_usernum_nlos_1010}
    \end{minipage}%
    \hfill
\begin{minipage}[t]{0.24\textwidth}
        \vspace{0pt}
        \centering
        \includegraphics[width=\linewidth]{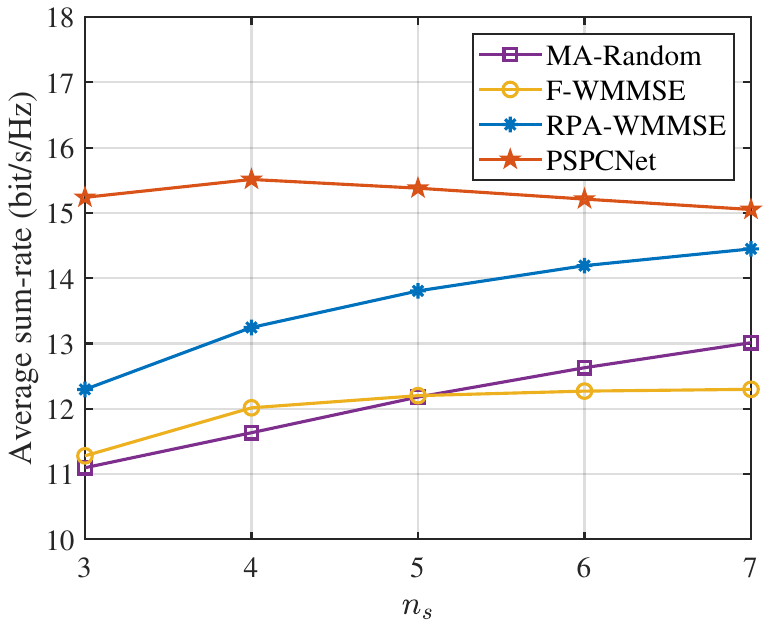}
        \caption{Average sum rate versus $n_{\rm s}$, $K=3$, NLOS, $\mathrm{SNR}=40$~dB, PSPCNet trained at $n_{\rm s}=4$.}
        \label{fig:sr_portnum_nlos_1010}
    \end{minipage}%
    \hfill
\begin{minipage}[t]{0.24\textwidth}
        \vspace{0pt}
        \centering
        \includegraphics[width=\linewidth]{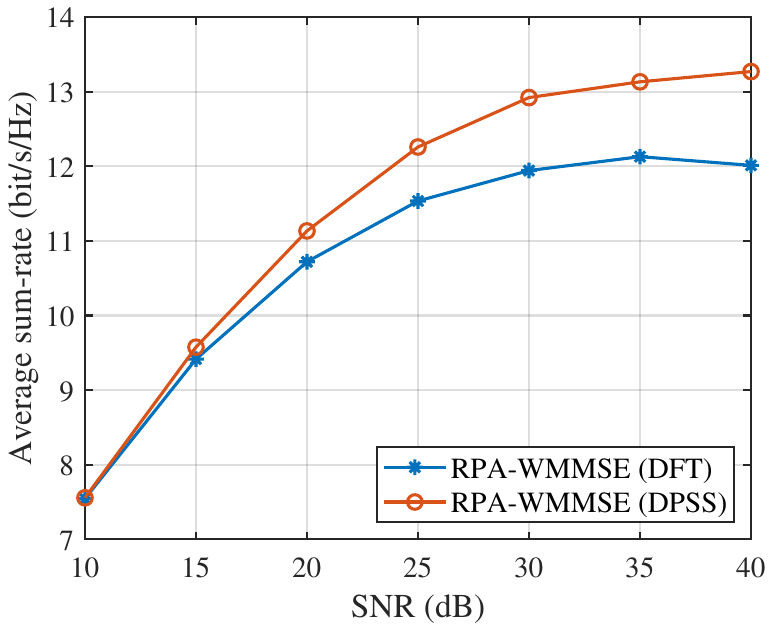}
        \caption{Average sum rate versus SNR for RPA-WMMSE using DFT and DPSS bases, $n_{\rm s}=4$, $K=3$, NLOS.}
        \label{fig:sr_snr_dft_dpss_nlos_1010}
    \end{minipage}%
    \hfill
\begin{minipage}[t]{0.24\textwidth}
        \vspace{0pt}
        \centering
        \includegraphics[width=\linewidth]{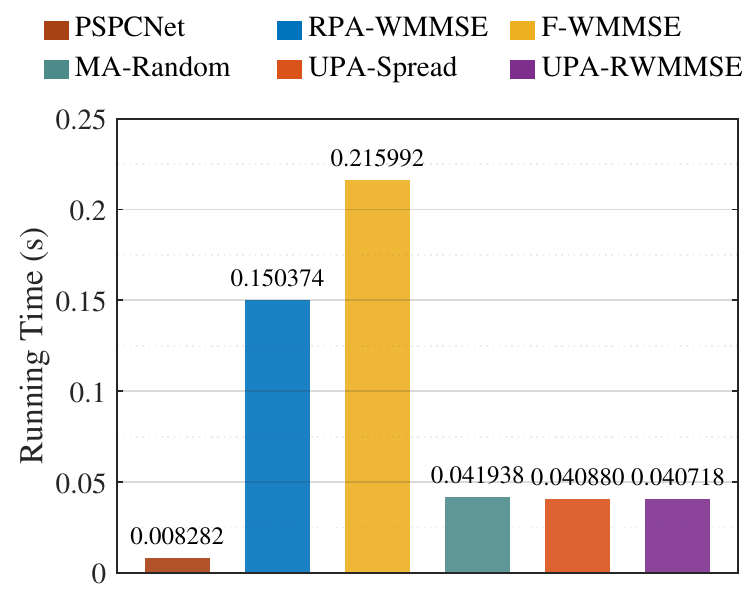}
        \caption{Average runtime comparison with a CSI delay of $10$~ms, $n_{\rm s}=4$, $K=3$, NLOS, $\mathrm{SNR}=30$~dB.}
        \label{fig:runtime_nlos_1010}
    \end{minipage}
    \vspace{-14pt}
\end{figure*}



\begin{figure*}[!t]
    \centering
\begin{minipage}[t]{0.24\textwidth}
    \vspace{0pt}
    \centering
    \includegraphics[width=\linewidth]{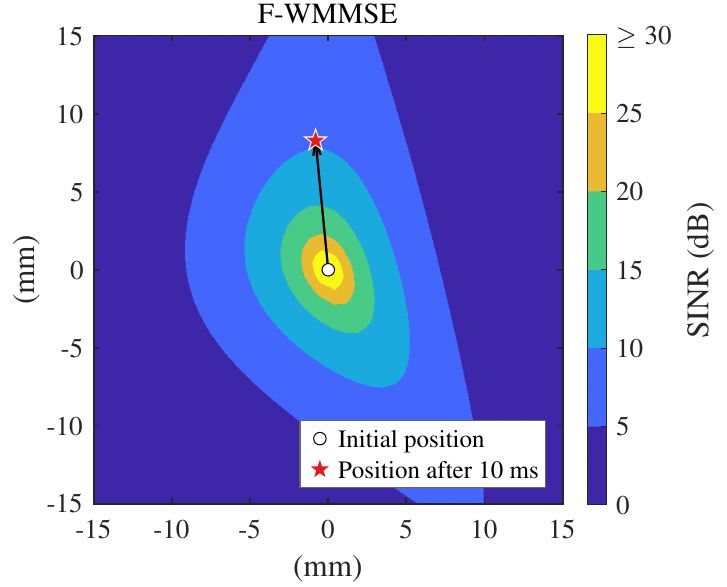}
    \caption{SINR distribution over the user movement region with F-WMMSE.}
    \label{fig:u3_spatial_sinr_fwmmse}
\end{minipage}%
    \hfill
\begin{minipage}[t]{0.24\textwidth}
    \vspace{0pt}
    \centering
    \includegraphics[width=\linewidth]{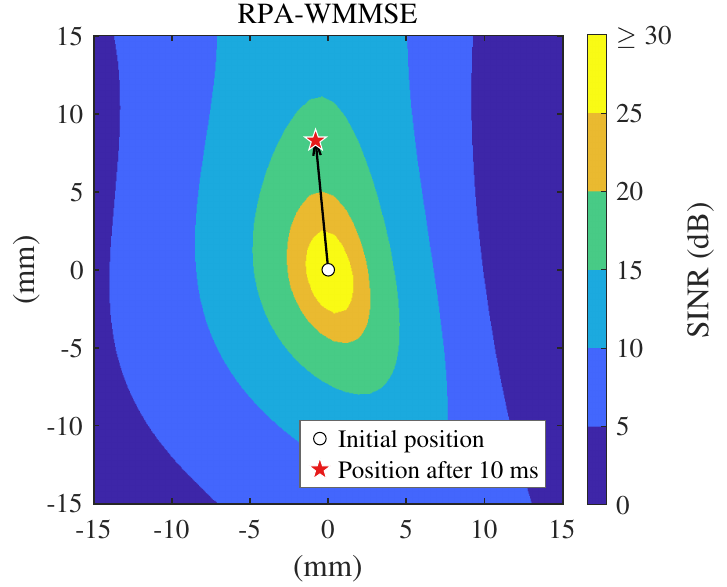}
    \caption{SINR distribution over the user movement region with RPA-WMMSE.}
    \label{fig:u3_spatial_sinr_rpa}
\end{minipage}%
    \hfill
\begin{minipage}[t]{0.24\textwidth}
        \vspace{0pt}
        \centering
        \includegraphics[width=\linewidth]{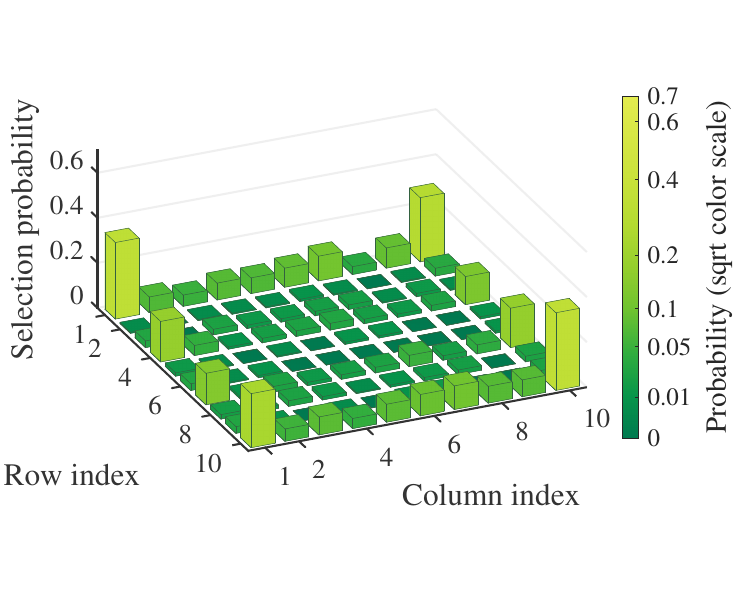}
        \caption{Position selection probabilities of RPA-WMMSE using the DPSS basis, $n_{\rm s}=4$, $K=3$, NLOS, $\mathrm{SNR}=0$~dB.}
        \label{fig:port_selection_dpss_0db}
    \end{minipage}%
    \hfill
\begin{minipage}[t]{0.24\textwidth}
        \vspace{0pt}
        \centering
        \includegraphics[width=\linewidth]{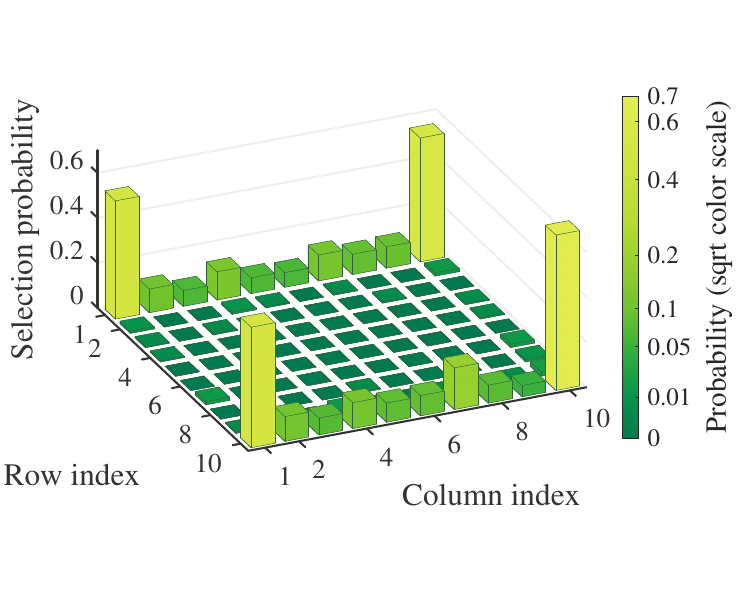}
        \caption{Position selection probabilities of RPA-WMMSE using the DPSS basis, $n_{\rm s}=4$, $K=3$, NLOS, $\mathrm{SNR}=40$~dB.}
        \label{fig:port_selection_dpss_40db}
    \end{minipage}
    \vspace{-14pt}
\end{figure*}

{\leavevmode\color{black}Figs.~\ref{fig:sr_usernum_nlos_1010} and~\ref{fig:sr_portnum_nlos_1010} examine the zero-shot generalization performance of PSPCNet across different numbers of users and MAs, respectively. In Fig.~\ref{fig:sr_usernum_nlos_1010}, the network is trained with $K=4$ and $n_{\rm s}=6$ and directly applied to configurations with $K=2,\ldots,6$. Across the tested user counts, PSPCNet maintains sum-rate performance comparable to that of RPA-WMMSE. In Fig.~\ref{fig:sr_portnum_nlos_1010}, the network is trained with $K=3$ and $n_{\rm s}=4$ and directly applied to \textcolor{black}{antenna positioning} and precoding with $n_{\rm s}=3,\ldots,7$, achieving higher sum rates than RPA-WMMSE, F-WMMSE, and MA-Random in all tested configurations. These results demonstrate that PSPCNet can generalize across different numbers of users and MAs without retraining while maintaining performance.}

To verify the performance advantage of the adopted DPSS basis, Fig.~\ref{fig:sr_snr_dft_dpss_nlos_1010} compares the average sum rates of RPA-WMMSE using DPSS and DFT. It can be observed that DPSS achieves a higher sum rate than DFT, with
the gain increasing with SNR and reaching $1.26$ bit/s/Hz
at $40$ dB. These results demonstrate the sum-rate benefit of DPSS-based
statistical channel modeling.

{\leavevmode\color{black}Fig.~\ref{fig:runtime_nlos_1010} compares the average
online execution times of the considered schemes.
The experiments are conducted using PyTorch~2.4.1 on a platform equipped with Intel Xeon Platinum 8336C CPU (2.30~GHz) and NVIDIA GeForce RTX 4090 GPU.
As shown in the figure, RPA-WMMSE offers a runtime
advantage over F-WMMSE, while PSPCNet runs
substantially faster than all other considered schemes.
Together with the preceding sum-rate comparisons,
these results demonstrate that the proposed learning
framework substantially reduces online execution time
while maintaining high sum-rate performance.}

\textcolor{black}{Figs.~\ref{fig:u3_spatial_sinr_fwmmse} and~\ref{fig:u3_spatial_sinr_rpa} compare the spatial SINR distributions under mobility-induced CSI errors in a two-user NLOS realization with $n_{\rm s}=4$ and $\mathrm{SNR}=40$~dB. A higher SINR corresponds to a higher achievable rate. F-WMMSE and RPA-WMMSE are obtained with a $10$-ms CSI delay. In this realization, F-WMMSE fails to maintain a high SINR as the user moves away from the initial position. In contrast, RPA-WMMSE exploits channel statistics to account for CSI uncertainty and maintains a broader high-SINR region, demonstrating the effectiveness of robust precoding.}
Figs.~\ref{fig:port_selection_dpss_0db} and~\ref{fig:port_selection_dpss_40db} show the distribution of optimized position selections. As the SNR increases from $0$ dB to $40$ dB, the selected positions, which are more frequently distributed along the aperture edges, become further concentrated at the four corners. However, considering the sum-rate curve of UPA-Spread in Fig.~\ref{fig:sr_snr_nlos_1010}, the scheme with fixed corner positions still performs worse than the proposed joint optimization method. This indicates that a higher selection probability at the corners does not mean that a fixed corner layout can fully exploit the transmission advantages of MA systems, optimizing the position combination based on channel information can still yield further performance gains.

\FloatBarrier

\section{Conclusion}\label{Section 7}
\textcolor{black}{This paper investigated robust joint \textcolor{black}{antenna positioning} and precoding for multiuser MA systems under imperfect CSI.} A DPSS-based \textit{a posteriori} channel model was established to characterize spatially correlated uncertainty. \textcolor{black}{We then proposed the RPA-WMMSE, which compares position sets through closed-form incremental scores, avoiding separate precoding iterations for each candidate.} \textcolor{black}{We further developed PSPCNet with user-interaction encoding, CSI-conditioned covariance attention, and distance-biased spatial attention to reduce the inference time.} \textcolor{black}{\textcolor{black}{Simulations demonstrate the benefits of posterior channel statistics and show that PSPCNet achieves high sum rates with substantially lower online runtime while supporting different numbers of users and MAs without retraining.}}

\appendices
\section{Proof of Proposition~\ref{dpss-review:prop:dpss}}\label{app:proof-dpss-reference}
Let $\mathbf a(f)=[1,e^{j2\pi f},\ldots,e^{j2\pi(N-1)f}]^T$ denote the spatial response at normalized frequency $f$. Under the reference model, the scattering components at distinct spatial frequencies are uncorrelated and their spatial power spectral density equals unity over $[-\delta,\delta]$. Their covariance contributions therefore add as
\begin{equation}
    \mathbf R_{\rm ref}
    =\int_{-\delta}^{\delta}\mathbf a(f)\mathbf a^H(f)\,\mathrm df.
    \label{eq:dpss-proof-covariance-integral}
\end{equation}
Evaluating the $(m,n)$-th entry gives $2\delta$ for $m=n$ and $\sin(2\pi\delta(m-n))/(\pi(m-n))$ otherwise, which yields~\eqref{dpss-review:eq:reference-covariance}.


\section{Proof of Proposition~\ref{prop:posterior-position-score-review}}\label{app:proof-position-score}
Let $\Delta\mathbf P=\mathbf P-\mathbf P(\mathcal S)$. Completing the square yields
\begin{equation}
    f(\mathbf E_{\mathcal S},\mathbf P)
    =\operatorname{Tr}\!\left(
    \Delta\mathbf P^{H}
    \mathbf W_{\mathcal S\mathcal S}
    \Delta\mathbf P
    \right)-\Gamma(\mathcal S).
    \label{eq:position-score-completing-square}
\end{equation}
Since $\mathbf W_{\mathcal S\mathcal S}\succ\mathbf0$, the trace term is nonnegative and equals zero if and only if $\Delta\mathbf P=\mathbf0$, which establishes the result.

\section{Proof of Proposition~\ref{prop:incremental-scoring-review}}\label{app:proof-incremental-score}
For nonempty $\mathcal T$, $\mathbf W\succ\mathbf0$ implies $\kappa_n(\mathcal T)>0$. Applying the block inverse identity~\cite{zhang2005schur} to~\eqref{eq:port-set-score} yields
\begin{equation}
\begin{aligned}
    &\Gamma(\mathcal T\cup\{n\})-\Gamma(\mathcal T)\\
    &=\frac{\left\|\mathbf u_n-\mathbf U_{\mathcal T}^{H}
    \mathbf W_{\mathcal T\mathcal T}^{-1}\mathbf w_n(\mathcal T)\right\|_2^2}
    {\kappa_n(\mathcal T)}\\
    &=\frac{\|\mathbf b_n(\mathcal T)\|_2^2}{\kappa_n(\mathcal T)},
\end{aligned}
    \label{eq:incremental-score-decomposition}
\end{equation}
where the last equality follows from~\eqref{eq:port-schur-residual}. Together with~\eqref{eq:incremental-score-definition}, this proves~\eqref{eq:port-schur-gain}. For $n_{\rm s}=1$, the result follows from $\Gamma(\varnothing)=0$ and $\Gamma(\{n\})=\|\mathbf u_n\|_2^2/W_{nn}$.

	\bibliographystyle{IEEEtran}
    \bibliography{FAS_refs}

\end{document}